\documentclass[11pt,a4paper,english]{article}
\pdfoutput=1

\usepackage{color}
\usepackage{mathtools}
\usepackage{amsmath}
\usepackage{amssymb}
\usepackage{graphicx}
\usepackage{esint}

\makeatletter

\newif\ifprstyle
\prstyletrue
\prstylefalse

\ifprstyle
  \usepackage[
    colorlinks=true, pdfa=true,
    urlcolor=blue,   anchorcolor=blue,
    citecolor=blue,  filecolor=blue,
    linkcolor=blue,  menucolor=blue,
    pagecolor=blue,  linktocpage=true
  ]{hyperref}
\else
  \usepackage{jheppub}
  \def\@fpheader{\\}
  
\fi

\usepackage{amsfonts}

\usepackage{tensind} 
\tensordelimiter{?}

\usepackage{xcolor}
\usepackage{colortbl}

\let\myHlineC\hline

\newcommand{\bHlineC}{
  \renewcommand{\hline}{\arrayrulecolor{lightgray}\myHlineC\arrayrulecolor{black}}
  \newcolumntype{|}{!{\color{lightgray}\vline}}
}

\newcommand{\eHlineC}{
  \renewcommand{\hline}{\myHlineC}
  \newcolumntype{|}{!{\color{black}\vline}}
}

\ifprstyle
  \newcommand{\repEq}[2]{\tag{\ref{#1}#2}}
\else
  \newcommand{\repEq}[2]{\tag{\ref*{#1}#2}}
\fi

\newcommand{\bSe}{\begin{subequations}} 
\newcommand{\eSe}{\end{subequations}}

\newcommand{\bWe}{\begin{widetext}} 
\newcommand{\eWe}{\end{widetext}}

\allowdisplaybreaks

\usepackage{tikz,amsmath,amssymb,bm,color,cancel}
\usetikzlibrary{shapes,arrows}
\usetikzlibrary{calc}
\usetikzlibrary{positioning}
\usetikzlibrary{decorations.pathreplacing}

\makeatother

\usepackage{babel}
\begin{document}


\global\long\def\dd{\mathrm{d}}
\global\long\def\ee{\mathrm{e}}
\global\long\def\ii{\mathrm{i}}
\global\long\def\diag{\operatorname{diag}}
\global\long\def\sign{\operatorname{sign}}
\global\long\def\sign{\operatorname{sign}}
\global\long\def\artanh{\operatorname{artanh}}
\global\long\def\Tr{\operatorname{Tr}}
\global\long\def\eH{\mathrm{{\scriptscriptstyle H}}}
\global\long\def\tr{\mathsf{{\scriptscriptstyle T}}}

\global\long\def\tudu#1#2#3#4{?{\mbox{\ensuremath{#1}}}^{#2}{}_{#3}{}^{#4}?}
\global\long\def\tdud#1#2#3#4{?{\mbox{\ensuremath{#1}}}{}_{#2}{}^{#3}{}_{#4}?}
\global\long\def\tud#1#2#3{?{\mbox{\ensuremath{#1}}}^{#2}{}_{#3}?}
\global\long\def\tdu#1#2#3{?{\mbox{\ensuremath{#1}}}{}_{#2}{}^{#3}?}


\newcommand{\myAbstract}{We consider the Hassan-Rosen bimetric field
equations in vacuum when the two metrics share a single common null
direction in a spherically symmetric configuration. By solving these
equations, we obtain a class of exact solutions of the generalized
Vaidya type parametrized by an arbitrary function.  Besides not
being asymptotically flat, the found solutions are nonstationary admitting
only three global spacelike Killing vector fields which are the generators
of spatial rotations. Hence, these are spherically symmetric bimetric
vacuum solutions with the minimal number of isometries. The absence
of staticity formally disproves an analogue statement to Birkhoff's
theorem in the ghost-free bimetric theory which would state that a
spherically symmetric solution is necessarily static in empty space.}

\newcommand{\myTitle}{On Birkhoff's theorem in ghost-free bimetric
theory}

\newcommand{\myKeywords}{Modified gravity, Ghost-free bimetric theory,
Birkhoff's theorem}

\title{\myTitle}

\author{Mikica Kocic,}
\author{Marcus H\"{o}g\r{a}s,}
\author{Francesco Torsello}
\author{and Edvard M\"{o}rtsell}

\affiliation{
  Department of Physics \& The Oskar Klein Centre,\\
  Stockholm University, AlbaNova University Centre,
  SE-106 91 Stockholm
}

\emailAdd{mikica.kocic@fysik.su.se}
\emailAdd{marcus.hogas@fysik.su.se}
\emailAdd{francesco.torsello@fysik.su.se}
\emailAdd{edvard@fysik.su.se}

\ifprstyle
  \begin{abstract}\myAbstract\end{abstract}
\else
  \abstract{\myAbstract}
\fi


\maketitle
\ifprstyle\else
  \flushbottom
\fi

\section{Introduction and summary}

The framework of this paper is the Hassan-Rosen (HR) ghost-free bimetric
theory \cite{Hassan:2011zd}, which is a classical nonlinear theory
of two interacting spin-2 fields. As shown in \cite{Hassan:2011zd,Hassan:2011ea},
the HR bimetric theory is free of instabilities such as the Boulware-Deser
ghost \cite{Boulware:1973my}. An un\-am\-big\-u\-ous definition
of the theory which guaranties the existence of a spacetime interpretation
is given in \cite{Hassan:2017ugh}. The HR theory is closely related
to de Rham-Gabadadze-Tolley (dRGT) massive gravity \cite{deRham:2010ik,deRham:2010kj},
which is a nonlinear theory of a massive spin-2 field, proven to be
ghost-free in \cite{Hassan:2011hr}. For recent reviews of these theories,
see \cite{Schmidt-May:2015vnx,deRham:2014zqa}. 

Although understanding of the HR theory has seen a considerable development
in recent years, not many exact vacuum solutions have been found.
This is not surprising as the bimetric field equations are more than
doubled in number when compared to General relativity (GR). Consequently,
the majority of known exact vacuum solutions have both metrics in
standard GR form (see \cite{Schmidt-May:2015vnx} and references
therein, and also more recently \cite{Sushkov:2015fma,Nersisyan:2015oha,Mazuet:2015pea,Babichev:2015xha,Enander:2015kda,Li:2016fbf,Hu:2016hpm,Li:2016tcn,Torsello:2017cmz}),
where the only nonstationary and spherically symmetric vacuum solution
was found in \cite{Hassan:2012gz}. The other issue is an analogue
statement to Birkhoff's theorem \cite{Birkhoff:1923,Eiesland:1925,Schleich:2009uj}
which would claim that spherically symmetric bimetric solutions in
empty space are necessarily static.\footnote{~The theorem was published by J.T.~Jebsen \cite{Jebsen:1921} two
years before Birkhoff (reprinted in \cite{Jebsen:2005}).} It is argued that such a statement is absent in the bimetric theory
\cite{Babichev:2015xha}.

The main result of this paper is a class of exact bimetric vacuum
solutions which are nonstationary where the null tetrads \cite{Stephani:2009exact}
of the two metrics have a single common null direction in a spherically
symmetric configuration. Contrary to GR, where spherically symmetric
vacuum solutions admit at least four isometries, the found solutions
have only three Killing vector fields that are the generators of spatial
rotations. Moreover, the solutions are conformally flat. The metrics
are of the generalized Vaidya type parametrized by an arbitrary function,
here denoted the curvature field. For a constant curvature field,
one gets a proportional and maximally symmetric GR solution. The nonstationarity
of this and the solution from \cite{Hassan:2012gz} formally contradicts
an analogue statement to Birkhoff's theorem for the ghost-free bimetric
theory. Before summarizing results in more detail, we overview the
ghost-free bimetric theory and its field equations in the absence
of matter.

\subsection{Ghost-free bimetric action and equations of motion}

\label{sec:background}

In vacuum, the Hassan-Rosen action comprises two Einstein-Hilbert
terms with Planck masses $M_{g}$ and $M_{f}$ coupled through the
ghost-free interaction term \cite{Hassan:2011zd},
\begin{align}
S_{\mathrm{HR}} & =\frac{1}{2}M_{g}^{2}\!\int\dd^{4}x\sqrt{-g}\,R_{g}+\frac{1}{2}M_{f}^{2}\!\int\dd^{4}x\sqrt{-f}\,R_{f}-m^{4}\!\int\dd^{4}x\sqrt{-g}\,V(S).\label{eq:hr-action}
\end{align}
The absence of ghosts is ensured by the potential $V(S)$ of the following
form,
\begin{equation}
V(S)\coloneqq\sum_{n=0}^{4}\beta_{n}\,e_{n}(S),\label{eq:V}
\end{equation}
where $S$ denotes the square root of the operator $g^{\mu\rho}f_{\rho\nu}$.
In matrix notation, $S$ is the square root matrix function $S=\sqrt{g^{-1}f}$.
The potential is parametrized by real constants, $\beta_{n}$, $n=0,...,4$,
which are free parameters of the theory. The coefficients $e_{n}(S)$
in (\ref{eq:V}) are the elementary symmetric polynomials, which are
the scalar invariants of $S$ obtained through the generating function
\cite{macdonald:1998a},\vspace{-0.5em}
\begin{equation}
E(t,S)=\det(I+tS)=\sum_{n=0}^{\infty}e_{n}(S)\,t^{n}.\label{eq:e-genf}
\end{equation}
Note that $e_{n}(S)=0$ for $n>4$ due to the Cayley-Hamilton theorem.

By varying (\ref{eq:hr-action}) with respect to $g$ and $f$, we
obtain the equations of motion \cite{Hassan:2014vja},\bSe\label{eq:hr-eom}
\begin{equation}
G_{g}{}^{\mu}{}_{\nu}+\frac{m^{4}}{M_{g}^{2}}V_{g}{}^{\mu}{}_{\nu}(S)=0,\qquad G_{f}{}^{\mu}{}_{\nu}+\frac{m^{4}}{M_{f}^{2}}V_{f}{}^{\mu}{}_{\nu}(S)=0.
\end{equation}
\eSe Here, $G_{g}$ and $G_{f}$ are the Einstein tensors of $g$
and $f$, respectively, given in operator form, while $V_{g}$ and
$V_{f}$ are contributions of the potential (\ref{eq:V}),\bSe\label{eq:hr-eff-V}
\begin{align}
V_{g}(S) & =\sum_{n=0}^{3}\beta_{n}\sum_{k=0}^{n}(-1)^{n+k}e_{k}(S)\,S^{n-k},\label{eq:Vg}\\
V_{f}(S) & =\sum_{n=0}^{3}\beta_{4-n}\sum_{k=0}^{n}(-1)^{n+k}e_{k}(S^{-1})\,S^{-n+k},\label{eq:Vf}
\end{align}
\eSe which are coupled through the algebraic identity \cite{Hassan:2014vja},
\begin{equation}
V_{g}(S)+\det(S)\,V_{f}(S)=V(S).\label{eq:V-identity}
\end{equation}
Finally, the equations of motion (\ref{eq:hr-eom}) are supplemented
by two Bianchi constraints,
\begin{equation}
\nabla_{\mu}\left[V_{g}{}^{\mu}{}_{\nu}(S)\right]=0,\qquad\tilde{\nabla}_{\mu}\left[V_{f}{}^{\mu}{}_{\nu}(S)\right]=0,\label{eq:bianchi}
\end{equation}
where $\nabla_{\mu}$ and $\tilde{\nabla}_{\mu}$ are the covariant
derivatives compatible with $g$ and $f$, respectively. However,
assuming a nonsingular $S$, the two Bianchi constraints are not independent
since the invariance of the interaction term under the diagonal diffeomorphism
group implies the identity \cite{Damour:2002ws},
\begin{equation}
\nabla_{\mu}\left[V_{g}{}^{\mu}{}_{\nu}(S)\right]+\det(S)\,\tilde{\nabla}_{\mu}\left[V_{f}{}^{\mu}{}_{\nu}(S)\right]=0.
\end{equation}

\subsection{Summary of results}

We consider bimetric field equations in vacuum when the null tetrads
of the two metrics share a single common null direction throughout
the spacetime in a spherically symmetric configuration. Locally, this
configuration is not simultaneously diagonalizable and referred to
as Type IIa by the algebraic classification of square roots in \cite{Hassan:2017ugh}.
By solving the equations in the spherically symmetric chart $(v,r,\theta,\phi)$,
we obtain the class of solutions parametrized by an arbitrary function
$\lambda(v)>0$,\bSe\label{eq:sol-intro}
\begin{align}
g & =-\left(1-\frac{1}{3}\Lambda(v)\,r^{2}\right)\dd v^{2}+2\,\dd v\,\dd r+r^{2}\left(\dd\theta^{2}+\sin^{2}\theta\,\dd\phi^{2}\right)\!,\\
f & =-\lambda^{2}(v)\left[\left(1-\frac{1}{3}\Lambda(v)\,r^{2}-2\frac{\lambda^{\prime}(v)}{\lambda(v)}\,r\right)\dd v^{2}+2\,\dd v\,\dd r+r^{2}\left(\dd\theta^{2}+\sin^{2}\theta\,\dd\phi^{2}\right)\right]\!,
\end{align}
\eSe where,
\begin{equation}
\Lambda(v)\coloneqq3\beta_{2}m^{4}\left(M_{f}^{-2}+M_{g}^{-2}\lambda^{2}(v)\right)\!.
\end{equation}
The field $\Lambda(v)$ completely defines the class of solutions
(\ref{eq:sol-intro}) through the parameters $\Lambda_{0}\coloneqq3\beta_{2}m^{4}M_{g}^{-2}$,
$\alpha\coloneqq M_{f}/M_{g}$, and the form of $\lambda(v)$. The
function $\lambda(v)$ is undetermined by equations of motion,  provided
that the values of the rest of the $\beta$-parameters  satisfy,
\begin{equation}
\beta_{0}=3\beta_{2}\alpha^{-2},\qquad\beta_{4}=3\beta_{2}\alpha^{2},\qquad\beta_{1}=\beta_{3}=0.\label{eq:pm-intro}
\end{equation}
These values are known as the partially massless (PM) parameters since
they provide a de Sitter background in the context of PM bimetric
gravity \cite{Hassan:2012gz}. For arbitrary $\beta$-parameters,
the equation of motion sets a constant $\lambda(v)$ with a value
given by the parameters. In the following, the class of solutions
(\ref{eq:sol-intro})\textendash (\ref{eq:pm-intro}) will also be
referred to simply as ``the solution''. 

The Weyl tensor vanishes identically in both sectors, so the solution
is conformally flat (Petrov Type O). As can be shown, the field $\Lambda(v)$
enters all curvature scalars; hence, the solution exhibits a variable
curvature parametrized by $\Lambda(v)$ which is accordingly called
the \emph{curvature field}. For a constant $\Lambda(v)$, the solution
becomes proportional and maximally symmetric (i.e., an ordinary GR
solution: Minkowski, de Sitter or anti-de Sitter). The solution (\ref{eq:sol-intro})
is not asymptotically flat, unless $\Lambda(v)=0$. 

Physically, the effective stress-energy tensors (\ref{eq:hr-eff-V})
of the solution are nonperfect null fluids of Type II in GR \cite{Hawking:1973large}
and the metrics can be classified as being of the generalized Vaidya
type. In GR, an ordinary Vaidya metric is a solution of the Einstein
field equations describing the spacetime of a spherically symmetric
inhomogeneous imploding (exploding) null dust fluid \cite{Vaidya:1951zza}.
However, the metrics (\ref{eq:sol-intro}) have no curvature singularities.
They have the form of the Husain null fluid spacetimes \cite{Husain:1995bf}
and the generalized Vaidya metric \cite{Wang:1998qx} analyzed in
\cite{Ibohal:2004kk,Ibohal:2006ez,Ibohal:2009px} in the context of
nonstationary de Sitter cosmological models in GR.

For a variable $\Lambda(v)$, the solution is nonstationary admitting
only three global spacelike Killing vector fields that are the generators
of spatial rotations. In GR, spherically symmetric vacuum solutions
have from ten (corresponding to de Sitter, Minkowski, and anti-de
Sitter metrics) to four Killing vector fields (the minimal symmetry)
\cite{Bokhari:1990a6}. Thus, contrary to GR, the found class of solutions
comprises only three Killing vector fields. Importantly, the absence
of staticity contradicts an analogue statement to Birkhoff's theorem
in the ghost-free bimetric theory. Nonetheless, the solution admits
conformal Killing vector fields (see subsection \ref{sec:isom} for
more details).

\begin{figure}
\noindent \begin{centering}
\hspace{-1mm}\includegraphics{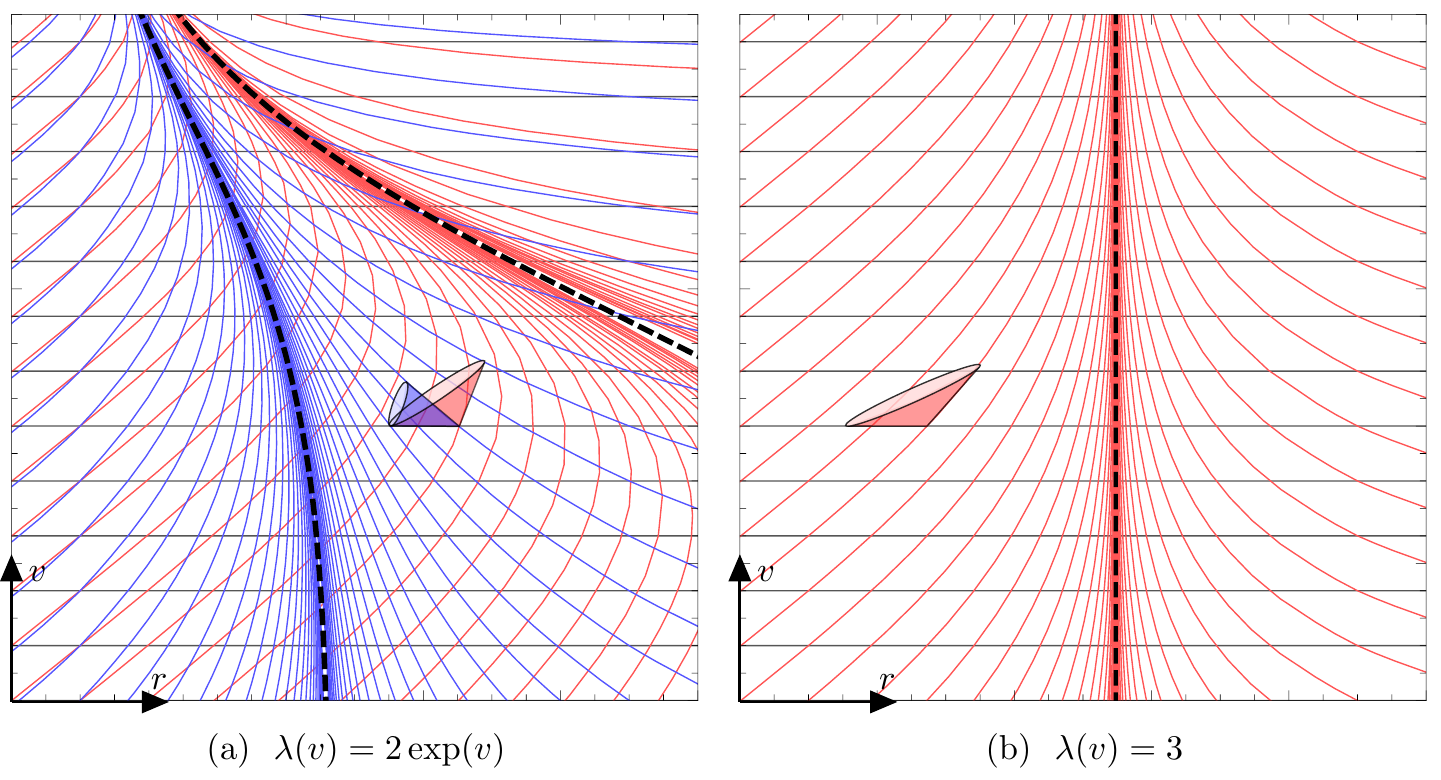}\vspace{-1mm}
\par\end{centering}
\caption{\label{fig:nc}Null radial geodesics for (a) $\lambda(v)=2\exp(v)$
and (b) $\lambda(v)=3$ for $\alpha=1$, $\beta_{2}=1/3$ with sample
future-pointing null cones. Ingoing null radial geodesics (common
to $g$ and $f$) are horizontal and depicted in black, outgoing geodesics
of $g$ are  in red and outgoing geodesics of $f$ are  in blue.
Dashed lines are the cosmological horizons. The geodesics of $g$
and $f$ coincide in the right panel since the solution is proportional.
Note the broken translational symmetry in the left panel.}
\end{figure}

The nonstationarity (and thus nonstaticity) of the solution is best
illustrated by plotting the radial null geodesics of $g$ and $f$
in a local patch $(v,r,\theta,\phi)$. This is done in Figure \ref{fig:nc}
for (a) the nonstationary case with $\lambda(v)\propto\exp(v)$, and
(b) the static case with $\lambda(v)=\mathrm{const}$. The broken
translational symmetry and the nonhomogeneity is clearly visible from
Figure \ref{fig:nc}(a). Also, the case (a) looks almost static when
$\lambda^{\prime}(v)\to0$ in the limit $v\to-\infty$.

The rest of this paper is organized as follows. The spacetime ansatz
and the equations of motion are stated in subsection \ref{sec:ansatz}.
The equations of motion are solved in subsection \ref{sec:eom}. The
properties of the found solution are given in subsection \ref{sec:sol-desc}.
The Killing and conformal Killing equations are solved in subsection
\ref{sec:isom}. A physical interpretation of the solution is given
in subsection \ref{sec:physics}. The paper ends with a short discussion
in section \ref{sec:discussion}. The relevant chart transition maps
are given in the appendix.

\section{Bimetric vacuum solutions of the generalized Vaidya type}

\label{sec:main}

\subsection{The configuration with one common null direction}

\label{sec:ansatz}

In this subsection, we write the ansatz for the spherically symmetric
bimetric spacetime where the two metrics share one common null direction,
denoted Type IIa in \cite{Hassan:2017ugh}.

A spherically symmetric metric is one which remains invariant under
rotations. In particular, the isometry group of a spherically symmetric
metric contains a subgroup isomorphic to SO(3). The orbits of this
subgroup are two-dimensional spheres. In a spherical chart $(\theta,\phi)$,
the metric on each orbit two-sphere is induced by the spacetime metric
and takes the form $r^{2}\left(\dd\theta^{2}+\sin^{2}\theta\,\dd\phi^{2}\right)$
where the scalar field $r^{2}$ parametrizes the area of the two-sphere
\cite{Wald:1984}. Completing the spherical chart with two additional
spacetime coordinates $a$ and $b$, the most general spherically
symmetric metric in a chart $(a,b,\theta,\phi)$ can be written,
\begin{equation}
-A(a,b)\,\dd a^{2}+B(a,b)\,\dd b^{2}+2C(a,b)\,\dd a\,\dd b+r(a,b)^{2}\dd\Omega^{2},
\end{equation}
where $\dd\Omega^{2}\coloneqq\dd\theta^{2}+\sin^{2}\theta\,\dd\phi^{2}$.
For two spherically symmetric metrics, we have two sets of fields
$A$, $B$, $C$ and $r$. The gauge freedom allows a reparametrization
of one of the coordinates, for example, setting $b=r$ along the radial
coordinate of one of the metrics. Since one of the metrics can always
be diagonalized, the bimetric spherically symmetric setup has six
independent scalar fields.

The second assumption is that the two metrics have a common null direction
throughout the spacetime. Suggestively choosing $a=v$, the ansatz
in the chart $(v,r,\theta,\phi)$ reads,\bSe\label{eq:ansatz}
\begin{align}
g & =-\ee^{2p(v,r)}G(v,r)\,\dd v^{2}+2\ee^{p(v,r)}\dd v\,\dd r+r^{2}\dd\Omega^{2},\label{eq:ansatz-g}\\
f & =\lambda^{2}(v,r)\,\left[-\ee^{2q(v,r)}F(v,r)\,\dd v^{2}+2\ee^{q(v,r)}\dd v\,\dd r+r^{2}\dd\Omega^{2}\right]\!,\label{eq:ansatz-f}
\end{align}
\eSe where $G$, $F$, $p$, $q$ and $\lambda$ are scalar fields.
Relative to both metrics, $v=\mathrm{const}$ are spherically symmetric
null surfaces. Compared to the most general ansatz, the absence of
the $\dd r^{2}$-component for $f$ is imposed by the common null
direction requirement. Further setting $p\equiv0$ and $q\equiv0$
makes the contributions of the potential (\ref{eq:hr-eff-V}) of Type
II in \cite{Hawking:1973large,Martin-Moruno:2017exc} (see also subsection
\ref{sec:physics}). As a consequence, the degrees of freedom in this
setup are contained in the metric fields $G(v,r)$, $F(v,r)$ and
$\lambda(v,r)$. At this point, there is no particular attribute attached
to the null coordinate $v$. After solving the equations, the coordinate
$v$ will be interpreted as the advanced (ingoing) time. Substituting
$v$ by $-u$, we can repeat our analysis and consider the chart in
terms of the retarded (outgoing) time $u$.

The complex null tetrads \cite{Stephani:2009exact} of $g$ and $f$
are given by the vector fields,\bSe\label{eq:null-tetrad}
\begin{alignat}{2}
\ell_{g} & =-\partial_{r}, & \quad\qquad\ell_{f} & =-\lambda(v,r)\,\partial_{r},\\
n_{g} & =\partial_{v}+\frac{1}{2}G(v,r)\,\partial_{r}, & n_{f} & =\lambda(v,r)\left(\partial_{v}+\frac{1}{2}F(v,r)\,\partial_{r}\right)\!,\\
m_{g} & =\frac{1}{\sqrt{2}r}\left(\partial_{\theta}+\ii\csc\theta\,\partial_{\phi}\right), & m_{f} & =\frac{1}{\sqrt{2}r}\lambda(v,r)\left(\partial_{\theta}+\ii\csc\theta\,\partial_{\phi}\right),
\end{alignat}
\eSe satisfying respective $\ell^{a}n_{a}=-1$, $m^{a}\bar{m}_{a}=1$
with all other contractions vanishing. Then, in component form, the
two metric inverses read,
\begin{equation}
g^{\mu\nu}=-2\ell_{g}^{(\mu}n_{g}^{\nu)}+2m_{g}^{(\mu}\bar{m}_{g}^{\nu)},\,\qquad f^{\mu\nu}=-2\ell_{f}^{(\mu}n_{f}^{\nu)}+2m_{f}^{(\mu}\bar{m}_{f}^{\nu)}.\qquad\ \ 
\end{equation}
Clearly, the null radial directions $\ell_{g}$ and $\ell_{f}$ are
proportional for the two metrics while the null radial directions
$n_{g}$ and $n_{f}$ become proportional for $F(v,r)=G(v,r)$. Also,
$m_{g}$ and $m_{f}$ are proportional as a result of spherical symmetry. 

\bHlineC

In matrix notation, the ansatz can be written,\bSe\label{eq:ansatz2}
\begin{align}
g & =\left(\begin{array}{cc|cc}
-G(v,r)\  & 1 & 0 & 0\\
1 & 0 & 0 & 0\\
\hline 0 & 0 & r^{2} & 0\\
0 & 0 & 0 & r^{2}\sin^{2}\theta
\end{array}\right)\!,\label{eq:mat-g}\\
f & =\left(\begin{array}{cc|cc}
-\lambda^{2}(v,r)F(v,r)\  & \lambda^{2}(v,r) & 0 & 0\\
\lambda^{2}(v,r) & 0 & 0 & 0\\
\hline 0 & 0 & \lambda^{2}(v,r)r^{2} & 0\\
0 & 0 & 0 & \lambda^{2}(v,r)r^{2}\sin^{2}\theta
\end{array}\right)\!.\label{eq:mat-f}
\end{align}
\eSe For $\lambda(v,r)>0$, the principal square root $S=\sqrt{g^{-1}f}$
reads,
\begin{equation}
S=\left(\begin{array}{cc|cc}
\lambda(v,r) & 0 & 0 & 0\\
\frac{1}{2}\lambda(v,r)\left(G(v,r)-F(v,r)\right)\  & \lambda(v,r) & 0 & 0\\
\hline 0 & 0 & \lambda(v,r) & 0\\
0 & 0 & 0 & \lambda(v,r)
\end{array}\right)\!.\label{eq:mat-S}
\end{equation}
In general, $S$ can have the Segre types $[1111]$, $[211]$, $[z\bar{z}11]$
and $[31]$. For $G(v,r)\ne F(v,r)$, the square root (\ref{eq:mat-S})
has Segre characteristics $[(211)]$, which is referred to as Type
IIa by the algebraic classification of bimetric solutions in \cite{Hassan:2017ugh}.
The contributions of the potential (\ref{eq:hr-eff-V}) are matrix-valued
functions of the square root (\ref{eq:mat-S}) and have the same Segre
type. Note that the stress-energy tensors of types $[31]$ and $[z\bar{z}11]$
are known to violate the weak energy condition in GR (see Theorem~7.11
in \cite{Hall:2004symmetries} and also \cite{Hall:1983a,Hawking:1973large}).
Hence, by GR standards, only types $[1111]$ and $[211]$ potentially
bear physical viability.

\eHlineC

\subsection{The equations of motion}

\label{sec:eom}

Here we present and solve the equations of motion for the ansatz (\ref{eq:ansatz2}).
For convenience, we introduce the Planck mass ratio $\alpha\coloneqq M_{f}/M_{g}$
and also absorb $m^{4}M_{g}^{-2}$ in the $\beta$-parameters, $m^{4}M_{g}^{-2}\beta_{n}\to\beta_{n}$.
We can always recover the mass terms by the reverse procedure at any
point. Note that the $\beta$-parameters are not any more dimensionless
after this substitution. The bimetric field equations (\ref{eq:hr-eom})
and the Bianchi constraint (\ref{eq:bianchi}) become,
\begin{align}
0 & =\tud{G_{g}}{\mu}{\nu}+\tud{V_{g}(S)}{\mu}{\nu},\label{eq:eom1}\\
0 & =\tud{G_{f}}{\mu}{\nu}+\alpha^{-2}\tud{V_{f}(S)}{\mu}{\nu},\label{eq:eom2}\\
0 & =\nabla_{\mu}\left[\tud{V_{g}(S)}{\mu}{\nu}\right].\label{eq:eom3}
\end{align}
To further simplify equations, following \cite{Kocic:2017wwf}, we
introduce the symmetric function $\left\langle \cdot\right\rangle _{k}^{n}$
for a set of variables $x_{1},\dots,x_{d}$ or a single repeated variable
$\lambda$, defined by,
\begin{equation}
\left\langle x_{1},\dots,x_{d}\right\rangle _{k}^{n}\coloneqq\sum_{i=0}^{n}\beta_{i+k}\,e_{i}(x_{1},\dots,x_{d}),\qquad\left\langle \lambda\right\rangle _{k}^{n}\coloneqq\sum_{i=0}^{n}\binom{n}{i}\beta_{i+k}\lambda^{i}.
\end{equation}
If the argument of $\left\langle \cdot\right\rangle _{k}^{n}$ is
a matrix, its eigenvalues are used; for example, we can express the
potential as $V(S)=\left\langle S\right\rangle _{0}^{4}$.


Substituting (\ref{eq:ansatz}) in (\ref{eq:eom1})-(\ref{eq:eom3})
gives the following system of equations,
\begin{align}
\repEq{eq:eom1}a0 & =r^{2}\left\langle \lambda\right\rangle _{0}^{3}+G-1+r\partial_{r}G,\label{eq:eom1a}\\
\repEq{eq:eom1}b0 & =r\lambda\left\langle \lambda\right\rangle _{1}^{2}(F-G)-2\partial_{v}G,\label{eq:eom1b}\\
\repEq{eq:eom1}c0 & =2r\left\langle \lambda\right\rangle _{0}^{3}+r\partial_{r}^{2}G+2\partial_{r}G,\label{eq:eom1c}\\
0 & =\alpha^{-2}r^{2}\lambda\left\langle \lambda\right\rangle _{1}^{3}+\lambda^{2}\left(F-1+r\partial_{r}F\right)+4r\lambda\left(F\partial_{r}\lambda+\partial_{v}\lambda\right)\nonumber \\
\repEq{eq:eom2}a & \qquad+\,r^{2}\partial_{r}\lambda\left(-F\partial_{r}\lambda+2\partial_{v}\lambda\right)+\lambda r^{2}\partial_{r}F\partial_{r}\lambda+2\lambda r^{2}\left(F\partial_{r}^{2}\lambda+\partial_{v}\partial_{r}\lambda\right),\label{eq:eom2a}\\
\repEq{eq:eom2}b0 & =2\left(\partial_{r}\lambda\right)^{2}-\lambda\partial_{r}^{2}\lambda,\label{eq:eom2b}\\
0 & =\alpha^{-2}r\lambda\left(F-G\right)\left\langle \lambda\right\rangle _{1}^{2}+2\lambda^{2}\partial_{v}F-8r\partial_{v}\lambda\left(F\partial_{r}\lambda+\partial_{v}\lambda\right)\nonumber \\
\repEq{eq:eom2}c & \qquad+\,2r\lambda\left(\partial_{v}F\partial_{r}\lambda-\partial_{r}F\partial_{v}\lambda+2F\partial_{v}\partial_{r}\lambda+2\partial_{v}^{2}\lambda\right),\label{eq:eom2c}\\
0 & =\alpha^{-2}r^{2}\lambda\left\langle \lambda\right\rangle _{1}^{3}+\lambda^{2}\left(F-1+r\partial_{r}F\right)+4r\lambda\left(F\partial_{r}\lambda+\partial_{v}\lambda\right)\nonumber \\
\repEq{eq:eom2}d & \qquad+\,r^{2}\partial_{r}\lambda\left(3F\partial_{r}\lambda+2\partial_{v}\lambda\right)+r^{2}\lambda\left(\partial_{r}F\partial_{r}\lambda+2\partial_{v}\partial_{r}\lambda\right),\label{eq:eom2d}\\
0 & =2\alpha^{-2}r\lambda\left\langle \lambda\right\rangle _{1}^{3}+\lambda^{2}\left(2\partial_{r}F+r\partial_{r}^{2}F\right)+4\lambda\left(F\partial_{r}\lambda+\partial_{v}\lambda\right)\nonumber \\
\repEq{eq:eom2}e & \qquad-\,2r\partial_{r}\lambda\left(F\partial_{r}\lambda+2\partial_{v}\lambda\right)+4r\lambda\left(\partial_{r}F\partial_{r}\lambda+F\partial_{r}^{2}\lambda+2\partial_{v}\partial_{r}\lambda\right),\label{eq:eom2e}\\
0 & =2r(F-G)\left\langle \lambda\right\rangle _{1}^{1}\partial_{r}\lambda-r\lambda\left\langle \lambda\right\rangle _{1}^{2}\left(\partial_{r}F-\partial_{r}G\right)\nonumber \\
\repEq{eq:eom3}a & \qquad-\,\left(F-G\right)\left\langle \lambda\right\rangle _{1}^{2}\left(2\lambda+3r\partial_{r}\lambda\right)-6r\left\langle \lambda\right\rangle _{1}^{2}\partial_{v}\lambda,\label{eq:eom3a}\\
\repEq{eq:eom3}b0 & =\left\langle \lambda\right\rangle _{1}^{2}\partial_{r}\lambda.\label{eq:eom3b}
\end{align}
Solving the system is done in several steps, summarized below. 

\paragraph*{Step 1.}

Since $\partial_{r}\lambda(v,r)=0$ from the Bianchi constraint (\ref{eq:eom3b}),
$\lambda(v,r)=\lambda(v)$.

\paragraph*{Step 2.}

From (\ref{eq:eom1a}) and (\ref{eq:eom1c}), we get $2(1-G)+r^{2}\partial_{r}^{2}G=0$.
Solving this yields,
\begin{equation}
G(v,r)=1+a(v)\,r^{2}+\frac{b(v)}{r},\label{eq:G1}
\end{equation}
where $a(v)$ and $b(v)$ are arbitrary functions.

\paragraph*{Step 3.}

After substituting (\ref{eq:G1}), from (\ref{eq:eom1a}) we can express
$a(v)$ in terms of $\lambda(v)$ as, 
\begin{equation}
a(v)=-\frac{1}{3}\left\langle \lambda(v)\right\rangle _{0}^{3}=-\frac{1}{3}\left(\beta_{0}+3\beta_{1}\lambda(v)+3\beta_{2}\lambda^{2}(v)+\beta_{3}\lambda^{3}(v)\right),\label{eq:a1}
\end{equation}
which we again substitute back in the equations. The simplification
of the derivatives of $a(v)$ is done using the relation, 
\begin{equation}
\partial_{v}\left\langle \lambda(v)\right\rangle _{k}^{n}=n\left\langle \lambda(v)\right\rangle _{k+1}^{n-1}\,\lambda^{\prime}(v)\quad\implies\quad a^{\prime}(v)=-\lambda^{\prime}(v)\left\langle \lambda(v)\right\rangle _{1}^{2}\,.
\end{equation}

\paragraph*{Step 4.}

From (\ref{eq:eom1b}), we can express $F(v,r)$ in terms of $\lambda(v)$
and $b(v)$ as,
\begin{align}
F(v,r) & =1-\frac{1}{3}r^{2}\left\langle \lambda(v)\right\rangle _{0}^{3}+\frac{b(v)}{r}-2r\frac{\lambda^{\prime}(v)}{\lambda(v)}+2\frac{b^{\prime}(v)}{\lambda r^{2}\left\langle \lambda(v)\right\rangle _{1}^{2}}\\
 & =G(v,r)-2r\frac{\lambda^{\prime}(v)}{\lambda(v)}+2\frac{b^{\prime}(v)}{\lambda r^{2}\left\langle \lambda(v)\right\rangle _{1}^{2}}\,.
\end{align}

\paragraph*{Step 5.}

Subtracting (\ref{eq:eom2a}) and (\ref{eq:eom2e}), we obtain $b^{\prime}(v)/\left\langle \lambda(v)\right\rangle _{1}^{2}=0$;
thus $b(v)=b_{0}$ where $b_{0}$ is a real constant.\label{par:step-5}

\paragraph*{Step 6.}

Using $G(v,r)=1-\frac{1}{3}r^{2}\left\langle \lambda(v)\right\rangle _{0}^{3}+\frac{b_{0}}{r}$
and $F(v,r)=1-\frac{1}{3}r^{2}\left\langle \lambda(v)\right\rangle _{0}^{3}+\frac{b_{0}}{r}-2r\frac{\lambda^{\prime}(v)}{\lambda(v)}$
significantly reduces the equations to become mostly algebraical;
for example, equation (\ref{eq:eom2a}) becomes $\alpha^{2}\lambda\left\langle \lambda\right\rangle _{0}^{3}-\left\langle \lambda\right\rangle _{1}^{3}=0$,
that is,
\begin{equation}
\alpha^{2}\beta_{3}\lambda^{4}(v)+\left(3\alpha^{2}\beta_{2}-\beta_{4}\right)\lambda^{3}(v)+3\left(\alpha^{2}\beta_{1}-\beta_{3}\right)\lambda^{2}(v)+\left(\alpha^{2}\beta_{0}-3\beta_{2}\right)\lambda(v)-\beta_{1}=0.\label{eq:PM-equation}
\end{equation}
For a given set of $\beta$-parameters, the above quartic equation
fully defines $\lambda(v)=\mathrm{const}$. 

Now, let us assume that $\lambda(v)$ is not constant, i.e., $\lambda^{\prime}(v)\ne0$.
Then (\ref{eq:PM-equation}) must hold for an arbitrary $\lambda(v)$.
The parameter combination that keeps $\lambda(v)$ undetermined is:
$\beta_{0}=3\beta_{2}\alpha^{-2}$, $\beta_{4}=3\beta_{2}\alpha^{2}$,
$\beta_{1}=\beta_{3}=0$. This gives,
\begin{equation}
a(v)=-\frac{1}{3}\left\langle \lambda(v)\right\rangle _{0}^{3}=-\beta_{2}\left(\alpha^{-2}+\lambda^{2}(v)\right).
\end{equation}
Note that $\lambda(v)$ can be chosen freely as an integration function
baring no dynamics.

\paragraph*{Step 7.}

Finally, using the above $\beta$-parameters reduces (\ref{eq:eom1})-(\ref{eq:eom3})
to a single equation $b_{0}\lambda^{\prime}(v)=0$, which requires
$b_{0}$ to vanish identically in order to have an arbitrary $\lambda(v)$.
Otherwise, if $\lambda(v)$ was taken to be constant in Step 6, $b_{0}$
would be arbitrary.\vspace{0.5em}

Collecting the results for $a(v)$ and $b(v)$ gives the final form
of $F(v,r)$ and $G(v,r)$ as a solution, which is presented in the
following subsection.

\subsection{The solution and its geometry}

\label{sec:sol-desc}

In this subsection, we quote the found solution and analyze its geometrical
properties.

By solving (\ref{eq:eom1})-(\ref{eq:eom3}), we obtained the class
of exact solutions,\bSe\label{eq:sol-GF}
\begin{align}
G(v,r) & =1-\beta_{2}\left(\alpha^{-2}+\lambda^{2}(v)\right)r^{2},\\
F(v,r) & =1-\beta_{2}\left(\alpha^{-2}+\lambda^{2}(v)\right)r^{2}-2\frac{\lambda^{\prime}(v)}{\lambda(v)}r,
\end{align}
\eSe where $\lambda(v)>0$ is\emph{ an arbitrary function} having
no dynamics. The dimensionfull $\beta_{2}$ and the dimensionless
$\alpha$ are the only free parameters of the theory. Solving the
equations of motion in one branch also gave,
\begin{equation}
\beta_{0}=3\beta_{2}\alpha^{-2},\qquad\beta_{4}=3\beta_{2}\alpha^{2},\qquad\beta_{1}=\beta_{3}=0.\label{eq:PM-cond}
\end{equation}
This condition is the same as the one imposed for the de Sitter background
in the context of partially massless bimetric gravity \cite{Hassan:2012gz}.

Using (\ref{eq:sol-GF}), the metrics (\ref{eq:ansatz}) become,\bSe\label{eq:sol}
\begin{align}
g & =-\left[1-\beta_{2}\left(\alpha^{-2}+\lambda^{2}(v)\right)r^{2}\right]\dd v^{2}+2\,\dd v\,\dd r+r^{2}\dd\Omega^{2},\label{eq:sol-g}\\
f & =\lambda^{2}(v)\left\{ -\left[1-\beta_{2}\left(\alpha^{-2}+\lambda^{2}(v)\right)r^{2}-2r\frac{\lambda^{\prime}(v)}{\lambda(v)}\right]\dd v^{2}+2\,\dd v\,\dd r+r^{2}\dd\Omega^{2}\right\} \!.\label{eq:sol-f}
\end{align}
\eSe The square root (\ref{eq:mat-S}) is accordingly,
\begin{equation}
S=r\lambda^{\prime}(v)\,\partial_{r}\otimes\dd v+\lambda(v)\left[\,\partial_{v}\otimes\dd v+\partial_{r}\otimes\dd r+\partial_{\theta}\otimes\dd\theta+\partial_{\phi}\otimes\dd\phi\,\right],\label{eq:sol-S}
\end{equation}
with the only off-diagonal component $\tud Srv=\frac{1}{2}\lambda\left(G-F\right)=r\,\lambda^{\prime}(v)$.

One of the features of the solution is a variable cosmological `constant',
or rather the \emph{curvature} \emph{field},
\begin{equation}
\Lambda(v)\coloneqq3\beta_{2}\left(\alpha^{-2}+\lambda^{2}(v)\right),\label{eq:cosmo-field}
\end{equation}
in terms of which we can write (\ref{eq:sol}) as,\bSe
\begin{align}
g & =-\left(1-\frac{1}{3}\Lambda(v)\,r^{2}\right)\dd v^{2}+2\,\dd v\,\dd r+r^{2}\dd\Omega^{2},\label{eq:sol-g1}\\
f & =-\lambda^{2}(v)\left[\left(1-\frac{1}{3}\Lambda(v)\,r^{2}-2\frac{\lambda^{\prime}(v)}{\lambda(v)}\,r\right)\dd v^{2}+2\,\dd v\,\dd r+r^{2}\dd\Omega^{2}\right]\!.\label{eq:sol-f1}
\end{align}
\eSe Equation (\ref{eq:cosmo-field}) completely defines the class
of solutions through the parameters $\beta_{2}$, $\alpha$ and the
form of $\lambda(v)$. Reinstating the original dimensionless $\beta_{2}$,
(\ref{eq:cosmo-field}) becomes $\Lambda(v)=3\beta_{2}m^{4}\left(M_{f}^{-2}+M_{g}^{-2}\lambda^{2}(v)\right)$.

In terms of the curvature field, the bimetric potential reads $V(S)=\alpha^{2}\Lambda^{2}(v)/(3\beta_{2})$.
The geometry of $g$ and $f$ is given by the Ricci and Kretschmann
scalars,
\begin{equation}
R_{g}=4\Lambda(v),\quad R_{f}=R_{g}\lambda^{-2}(v),\quad K_{g}=\frac{8}{3}\Lambda^{2}(v),\quad K_{f}=K_{g}\lambda^{-4}(v).
\end{equation}
Clearly, the solution exhibit variable curvature sourced by the curvature
field $\Lambda(v)$ (hence the name). The curvature scalars are all
finite for the allowed range of $\lambda(v)>0$. In the case of a
singular square root, $\lambda(v)=0$ introduces a singularity in
the $f$-sector. This is in accordance with the proposition from \cite{Torsello:2017cmz}. 

The Weyl tensor vanishes identically in both sectors, so the solution
is everywhere of Type O by Petrov classification \cite{Petrov:1969einstein}.
Subsequently, both sectors are conformally flat where gravitational
effects are due to the field energy of $\Lambda(v)$. Physically,
the effective stress-energy tensors (\ref{eq:hr-eff-V}) of the solution
can be interpreted as an inhomogeneous nonperfect null fluid of Type
II \cite{Hawking:1973large} with a nonvanishing energy flux and a
negative pressure (see subsection \ref{sec:physics} for more details).

For a constant $\lambda(v)\coloneqq c$, we have $G(v,r)=F(v,r)$
and the solution become proportional $f=c^{2}g$ of Type I since $\lambda^{\prime}(v)=0$
makes possible a diagonalization of the square root. This is a maximally
symmetric bi-Einstein solution with constant curvature (Minkowski,
de Sitter or anti-de Sitter, depending on $\Lambda$).

By accordingly adjusting $\alpha$ and a dimensionful $\beta_{2}$,
one can obtain the GR limit, $\alpha\to0$, and the massive gravity
limit, $\alpha\to\infty$ \cite{Schmidt-May:2015vnx}. In both limits,
$V(S)$ is constant imposing $\lambda=\mathrm{const}$ and a constant
curvature field. A similar behavior is obtained by letting $\lambda^{\prime}(v)\to0$.

Finally, we address the presence of the radially subleading term $2r\lambda^{\prime}(v)/\lambda(v)$
in the $f$ metric. As we shall see, this term can be \emph{moved}
from $f$ to $g$ by a suitable chart transition map which puts $f$
in the same form as $g$. (The origin of this term is the off-diagonal
component in the Jordan normal form of the square root matrix, so
it can never be eliminated by a similarity transformation.) Consider
the chart transition map from $(v,r,\theta,\phi)$ to $(V,R,\theta,\phi)$
defined by,\bSe\label{eq:ctmap1}
\begin{alignat}{2}
R(v,r) & =r\,\lambda(v)>0, & \quad\qquad\dd R & =\lambda(v)\,\dd r+r\lambda^{\prime}(v)\,\dd v,\label{eq:ctmap1a}\\
\lambda(v) & =\dd V(v)/\dd v>0, & \qquad\dd V & =\lambda(v)\,\dd v,\label{eq:ctmap1b}
\end{alignat}
\eSe with the nonvanishing Jacobian determinant $\lambda^{2}(v)$.
Noticing that,
\begin{equation}
r^{2}\left(\alpha^{-2}+\lambda^{2}(v)\right)=\alpha^{-2}r^{2}+R^{2}(v,r),
\end{equation}
after some algebra we obtain,
\begin{align}
f & =-\lambda^{2}(v)\left[1-\beta_{2}\left(\alpha^{-2}r^{2}+R^{2}\right)-2r\frac{\lambda^{\prime}(v)}{\lambda(v)}\right]\dd v^{2}\\
 & \qquad\qquad+\,2\lambda^{2}(v)\,\dd v\,\dd r+\lambda^{2}(v)r^{2}\dd\Omega^{2}\\
 & =-\left[1-\beta_{2}\left(\alpha^{-2}r^{2}+R^{2}\right)\right]\dd V^{2}+2\,\dd V\,\dd R+R^{2}\dd\Omega^{2}\\
 & =-\left(1-\frac{1}{3}\bar{\Lambda}(V)R^{2}\right)\dd V^{2}+2\,\dd V\,\dd R+R^{2}\dd\Omega^{2},
\end{align}
where we defined the following variables to show the similarity with
$g$,
\begin{equation}
\bar{\Lambda}(V)\coloneqq3\bar{\beta}_{2}\left(\bar{\alpha}^{-2}+\bar{\lambda}^{2}(V)\right),\quad\bar{\lambda}(V)\coloneqq\lambda(v)^{-1},\quad\bar{\beta}_{2}\coloneqq\beta_{2}\alpha^{-2}\quad\text{and}\quad\bar{\alpha}\coloneqq\alpha^{-1}.\label{eq:redef-gf}
\end{equation}
Thus, under the redefinition (\ref{eq:redef-gf}), $f$ in the chart
$(V,R,\theta,\phi)$ has the same form as $g$ in the chart $(v,r,\theta,\phi)$.
Importantly, this relation makes the isometries in the two sectors
to be of the same kind with the Killing vector fields related by (\ref{eq:ctmap1}),
see \cite{Torsello:2017zz}.

\subsection{Isometries of the solution}

\label{sec:isom}

In this subsection, we solve the Killing equation and find the isometries
of the solution. We deduce that the solution only admits three Killing
vector fields that are generators of spatial rotations. We also solve
the conformal Killing equation and find a conformal Killing vector
field which becomes the generator of staticity when $\lambda(v)$
becomes constant.

The solution (\ref{eq:sol-GF}) is spherically symmetric by construction.
It is easy to verify that the following standard SO(3) Killing vector
fields are the isometries of both $g$ and $f$,
\begin{equation}
L_{1}=\partial_{\phi},\quad L_{2}=-\sin\phi\,\partial_{\theta}-\cot\theta\,\cos\phi\,\partial_{\phi},\quad L_{3}=\cos\phi\,\partial_{\theta}-\cot\theta\,\sin\phi\,\partial_{\phi}\,,
\end{equation}
that is, $\mathcal{L}_{\xi}g=0=\mathcal{L}_{\xi}f$ for $\xi\in\{L_{1},L_{2},L_{3}\}$.

As noted earlier, the chart transition map (\ref{eq:ctmap1}) relates
two sectors so that any Killing vector field found in one sector can
be mapped into another. Therefore, without loss of generality, we
can consider only the isometries of the $g$-sector.

Before solving the general Killing equation, we find a possible static
vector field orthogonal to two-spheres given by hypersurfaces of $(\theta,\phi)$
at constant $(v,r)$. This is done by introducing a vector field $\xi$
which depends only on the $(v,r)$ coordinates,
\begin{equation}
\xi=\xi^{\mu}\partial_{\mu}=\xi^{0}(v,r)\,\partial_{v}+\xi^{1}(v,r)\,\partial_{r}+\xi^{2}(v,r)\,\partial_{\theta}+\xi^{3}(v,r)\,\partial_{\phi}\,.\label{eq:xi-1}
\end{equation}
For such $\xi$, the Killing equation $\mathcal{L}_{\xi}g=0$ reads,\bSe\label{eq:killing-1}
\begin{align}
0 & =-\partial_{v}\xi^{0}+\partial_{v}\xi^{1}+\beta_{2}r\left(\alpha^{-2}+\lambda^{2}\right)\xi^{1}+\beta_{2}r^{2}\left(\alpha^{-2}+\lambda^{2}\right)\partial_{v}\xi^{0}+\beta_{2}r^{2}\xi^{0}\lambda\lambda^{\prime},\\
0 & =-\partial_{r}\xi^{0}+\partial_{r}\xi^{1}+\partial_{v}\xi^{0}+\beta_{2}r^{2}\left(\alpha^{-2}+\lambda^{2}\right)\partial_{r}\xi^{0},\\
0 & =r\cos\theta\,\xi^{2}+\sin\theta\,\xi^{1},\\
0 & =\partial_{r}\xi^{0}=\xi^{1}=\partial_{v}\xi^{2}=\partial_{r}\xi^{2}=\partial_{v}\xi^{3}=\partial_{r}\xi^{3}.\label{eq:le-1}
\end{align}
\eSe We immediately obtain $\xi^{1}=0$ and $\xi^{2}=\mathrm{const}$,
$\xi^{3}=\mathrm{const}$. Then $\xi^{1}=0$ together with (\ref{eq:le-1})
sets $\xi^{2}=0$. Substituting these, we get,
\begin{align}
0 & =-\partial_{v}\xi^{0}+\beta_{2}r^{2}\left(\alpha^{-2}+\lambda^{2}\right)\partial_{v}\xi^{0}+\beta_{2}r^{2}\lambda\lambda^{\prime},\\
0 & =\partial_{v}\xi^{0},
\end{align}
which together with $\partial_{r}\xi^{0}=0$ implies a constant $\xi^{0}$.
Thus the equations reduce to $\beta_{2}r^{2}\lambda(v)\lambda^{\prime}(v)=0$.
Hence, the Killing vector field (\ref{eq:xi-1}) is possible only
for a maximally symmetric solution with $\lambda(v)=\mathrm{const}$,
in which case $\xi^{0}=\mathrm{const}$, $\xi^{1}=0$, $\xi^{2}=0$
and $\xi^{3}=\mathrm{const}$, i.e., the Killing vector field is a
linear combination of $\partial_{v}$ and $\partial_{\phi}$ (where
$\partial_{v}$ generates staticity).

Next we solve the Killing equation for a vector field $\xi$ which
depends on all coordinates,
\begin{equation}
\xi=\xi^{0}(v,r,\theta,\phi)\,\partial_{v}+\xi^{1}(v,r,\theta,\phi)\,\partial_{r}+\xi^{2}(v,r,\theta,\phi)\,\partial_{\theta}+\xi^{3}(v,r,\theta,\phi)\,\partial_{\phi}.\label{eq:xi-2}
\end{equation}
Because of the spherical symmetry, we can always align the coordinate
system so that $\dd\phi=0$, also setting $\xi^{3}=0$. The Killing
equation for $g$ with respect to (\ref{eq:xi-2}) reads,\bSe\label{eq:killing-2}
\begin{align}
0 & =-\partial_{v}\xi^{0}+\partial_{v}\xi^{1}+\beta_{2}r\left(\alpha^{-2}+\lambda^{2}\right)\xi^{1}+\beta_{2}r^{2}\left(\alpha^{-2}+\lambda^{2}\right)\partial_{v}\xi^{0}+\beta_{2}r^{2}\xi^{0}\lambda\lambda^{\prime},\\
0 & =-\partial_{r}\xi^{0}+\partial_{r}\xi^{1}+\partial_{v}\xi^{0}+\beta_{2}r^{2}\left(\alpha^{-2}+\lambda^{2}\right)\partial_{r}\xi^{0},\\
0 & =-\partial_{\theta}\xi^{0}+\partial_{\theta}\xi^{1}+r^{2}\partial_{v}\xi^{2}+\beta_{2}r^{2}\left(\alpha^{-2}+\lambda^{2}\right)\partial_{\theta}\xi^{0},\\
0 & =-\partial_{\phi}\xi^{0}+\partial_{\phi}\xi^{1}+\beta_{2}r^{2}\left(\alpha^{-2}+\lambda^{2}\right)\partial_{\phi}\xi^{0},\\
0 & =\partial_{\theta}\xi^{0}+r^{2}\partial_{r}\xi^{2},\\
0 & =\xi^{1}+r\partial_{\theta}\xi^{2},\\
0 & =r\cos\theta\,\xi^{2}+\sin\theta\,\xi^{1},\\
0 & =\partial_{r}\xi^{0}=\partial_{\phi}\xi^{0}=\partial_{\phi}\xi^{2}.
\end{align}
\eSe Clearly, $\xi^{0}$ does not depend on $r$ and $\phi$, and
$\xi^{2}$ does not depend on $\phi$; consequently $\partial_{\phi}\xi^{1}=0$
so that $\xi^{1}$ does not depend on $\phi$. Moreover, we can express,
\begin{equation}
\xi^{1}=-r\cot\theta\,\xi^{2},\qquad\partial_{r}\xi^{2}=-r^{-2}\partial_{\theta}\xi^{0},\qquad\partial_{\theta}\xi^{2}=\cot\theta\,\xi^{2}.
\end{equation}
Substituting gives,\bSe
\begin{align}
0 & =-\partial_{v}\xi^{0}-r\cot\theta\,\partial_{v}\xi^{2}+\beta_{2}r^{2}\left(\alpha^{-2}+\lambda^{2}\right)\left(\partial_{v}\xi^{0}-\cot\theta\,\xi^{2}\right)+\beta_{2}r^{2}\xi^{0}\lambda^{\prime},\\
0 & =-\xi^{2}+\partial_{v}\xi^{0}+r^{-1}\partial_{\theta}\xi^{0},\\
0 & =r\xi^{2}+r^{2}\partial_{v}\xi^{2}+\left(-1+\beta_{2}r^{2}\left(\alpha^{-2}+\lambda^{2}\right)\right)\partial_{\theta}\xi^{0}.
\end{align}
\eSe From the first equation we can solve for $\partial_{v}\xi^{2}$,
which will simplify the last equation into,
\begin{equation}
-\cot\theta\,\xi^{2}+r^{-1}\cot\theta\,\partial_{\theta}\xi^{0}+\partial_{v}\xi^{0}=0.
\end{equation}
Expressing $\partial_{v}\xi^{0}=r\left(\xi^{2}-\tan\theta\,\partial_{\theta}\xi^{0}\right)$,
then substituting back in the first equation gives,
\begin{equation}
\beta_{2}r^{2}\xi^{0}(v,\theta)\lambda(v)\lambda^{\prime}(v)=0.
\end{equation}
For an arbitrary $\lambda(v)$, this equation requires $\xi^{0}=0$.
Substituted back gives $\xi^{2}=0$ and $\xi^{1}=-r^{-1}\tan\theta\,\xi^{2}=0$.
Thus, all the components of $\xi$ are necessarily 0 iff $\mathcal{L}_{\xi}g=0$,
so there are no other Killing vector fields except those in SO(3). 

Nonetheless, the solution may have conformal Killing vector fields,
so we endeavor in solving the conformal Killing equation $\mathcal{L}_{\xi}g=2\chi g$
where $\chi$ is a scalar field. Using the ansatz (\ref{eq:xi-1})
for $\xi$, the conformal Killing equation is slightly more complicated
than (\ref{eq:killing-1}),\bSe\label{eq:killing-3}
\begin{align}
0 & =\left[1-\beta_{2}\left(\alpha^{-2}+\lambda^{2}\right)r^{2}\right]\chi-\left[1-\beta_{2}\left(\alpha^{-2}+\lambda^{2}\right)r^{2}\right]\partial_{v}\xi^{0}\nonumber \\
 & \qquad+\,\partial_{v}\xi^{1}+\beta_{2}\left(\alpha^{-2}+\lambda^{2}\right)r\xi^{1}+\beta_{2}r^{2}\xi^{0}\lambda\lambda^{\prime},\\
0 & =-2\chi+\partial_{r}\xi^{1}+\partial_{v}\xi^{0}-\left[1-\beta_{2}\left(\alpha^{-2}+\lambda^{2}\right)r^{2}\right]\partial_{r}\xi^{0},\\
0 & =\xi^{1}-r\chi,\\
0 & =r\cos\theta\,\xi^{2}+\sin\theta\,\xi^{1}-r\sin\theta\,\chi,\\
0 & =\partial_{r}\xi^{0}=\partial_{v}\xi^{2}=\partial_{r}\xi^{2}=\partial_{v}\xi^{3}=\partial_{r}\xi^{3}.
\end{align}
\eSe Besides $\xi^{1}(v,r)=r\,\chi(v,r)$, we conclude that $\xi^{2}=c_{1}$
and $\xi^{3}=c_{2}$ where $c_{1}$, $c_{2}$ are constants. Because
of the spherical symmetry, we can set $\xi^{3}=0$. Also, $\partial_{r}\xi^{0}=0$
implies that $\xi^{0}(v,r)$ does not depend on $r$. Consequently,
we denote $\xi^{0}(v,r)=V(v)$, and (\ref{eq:killing-3}) reduces
to,\bSe
\begin{align}
0 & =\chi+r\partial_{v}\chi-\left[1-\beta_{2}r^{2}\left(\alpha^{-2}+\lambda^{2}\right)\right]V^{\prime}+\beta_{2}r^{2}V\lambda\lambda^{\prime},\label{eq:chi-1}\\
0 & =-\chi+r\partial_{r}\chi+V^{\prime},\label{eq:chi-2}\\
0 & =c_{1}r^{2}\sin\theta.\label{eq:chi-3}
\end{align}
\eSe From (\ref{eq:chi-3}) we get $c_{1}=0$, thus, $\xi^{2}=0$.
Solving (\ref{eq:chi-2}) yields $\chi(v,r)=r\,X(v)+V^{\prime}(v)$
where $X(v)$ is an arbitrary function depending only on $v$. Substituting
this $\chi(v,r)$ in (\ref{eq:chi-1}) yields,
\begin{equation}
r\left[X(v)+V^{\prime\prime}(v)\right]+\frac{1}{3}r^{2}\left[6X^{\prime}(v)+2\Lambda(v)V^{\prime}(v)+\Lambda^{\prime}(v)V(v)\right]=0.
\end{equation}
The above equation must hold for any $r$. Therefore, $X(v)=-V^{\prime\prime}(v)$
and,
\begin{equation}
V^{\prime\prime\prime}(v)-\frac{1}{3}\Lambda(v)V^{\prime}(v)-\frac{1}{6}\Lambda^{\prime}(v)V(v)=0.\label{eq:ckvf-V}
\end{equation}
This is a third-order homogeneous linear differential equation, for
which the Wronskian is identically one. For $\Lambda(v)\ne\mathrm{const}$,
the equation (\ref{eq:ckvf-V}) has three linearly independent solutions
resulting in three proper conformal Killing vector fields. The algebraic
structure of $V(v)$ is determined by the form of $\Lambda(v)$. For
example, if $\Lambda(v)$ is a polynomial in $v$, the solution is
a holonomic function. 

The resulting conformal Killing vector field reads,
\begin{equation}
\xi=V(v)\,\partial_{v}+r\left(V^{\prime}(v)-rV^{\prime\prime}(v)\right)\partial_{r},\label{eq:ckvf}
\end{equation}
where also $\chi(v,r)=V^{\prime}(v)-rV^{\prime\prime}(v)$, so that
$\mathcal{L}_{g}\xi=2\chi\xi$. For $\Lambda(v)=\mathrm{const}$,
we have $V(v)=1$, $\xi=\partial_{v}$ and $\chi=0$. Note that the
three independent solutions of (\ref{eq:ckvf-V}) give rise to nine
conformal Killing vectors because of the spherical symmetry.

\begin{figure}
\noindent \begin{centering}
\hspace{-1mm}\includegraphics{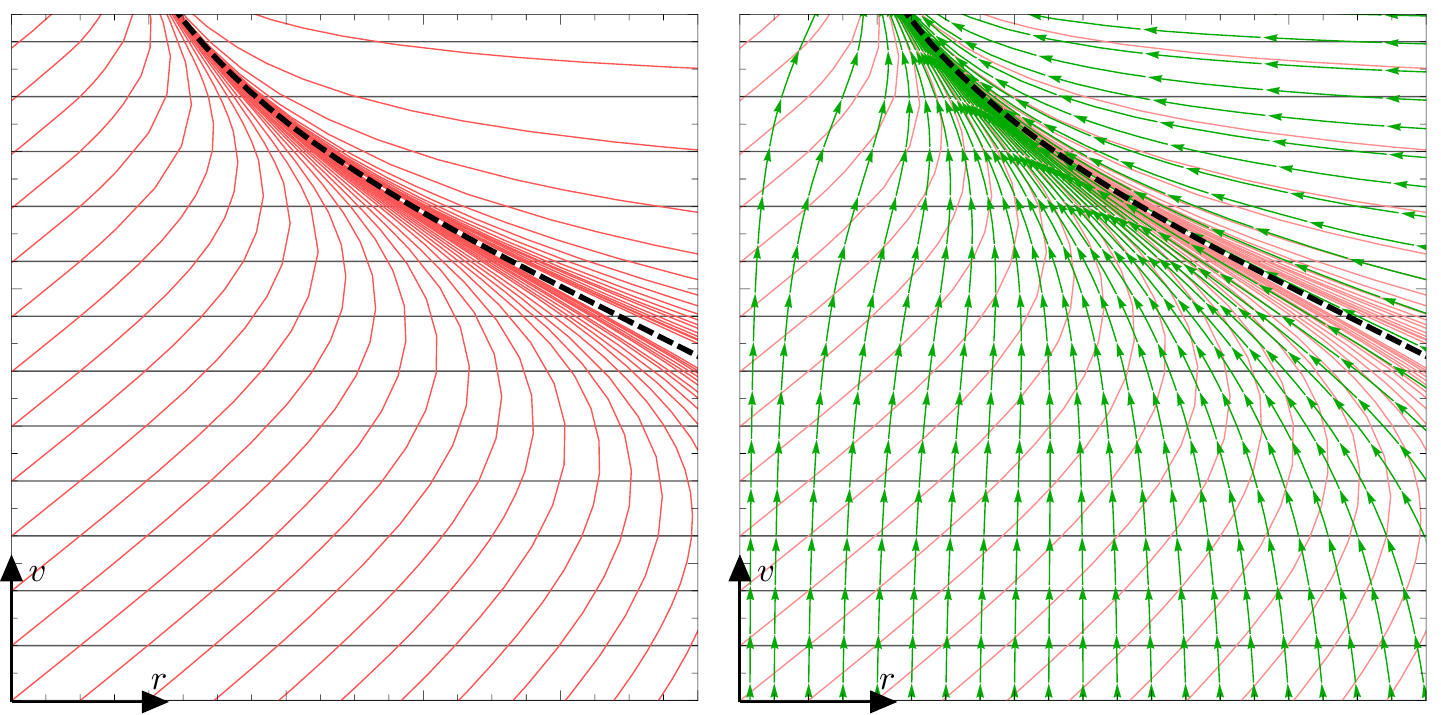}\vspace{-1mm}
\par\end{centering}
\caption{\label{fig:orbits}Orbits of the conformal Killing vector field $\xi$
(in green) covering the null radial geodesics of $g$ for $\lambda(v)=2\exp(v)$.
The ingoing and outgoing geodesics of $g$ are shown in black and
red, respectively. At $v\to-\infty$, the orbits become translational
along $\partial_{v}$.}
\end{figure}

As an illustration, let us revisit the example from Figure \ref{fig:nc}
where $\lambda(v)=2\exp(v)$. In such a case, one of the solutions
to (\ref{eq:ckvf-V}) reads,
\begin{equation}
V(v)={_{1}}F{_{2}}\left(\,\frac{1}{2}\,;\,1-\frac{\sqrt{\beta_{2}}}{2\alpha},\,1+\frac{\sqrt{\beta_{2}}}{2\alpha}\,;\,\beta_{2}\ee^{2v}\right)\!,\label{eq:ckf-V-ex}
\end{equation}
where $_{p}F_{q}$ denotes the generalized hypergeometric function.
The orbits of (\ref{eq:ckvf}) for $V(v)$ from (\ref{eq:ckf-V-ex})
are plotted in Figure \ref{fig:orbits}. In the constant curvature
limit $\lambda^{\prime}(v)\to0$, we have $V(v)\to1$ and $V^{\prime}(v)\to0$;
thus, the conformal Killing vector field (\ref{eq:ckvf}) reduces
to a timelike Killing vector field $\xi=\partial_{v}$ becoming the
generator of staticity.

\subsection{Physical interpretation}

\label{sec:physics}

In this subsection, we first argue why Birkhoff's theorem cannot be
stated in bimetric theory in general, and then give a physical interpretation
of the found solution.

We know that staticity is recovered when the curvature field $\Lambda(v)$
flattens to a constant. In such a case, $\Lambda(v)$ becomes an ordinary
cosmological constant and the two metrics become dynamically decoupled
forming a bi-Einstein solution (all the terms in (\ref{eq:V-identity})
are constant in this case). Consequently, both metrics behave as they
have separate vacua, so the extended version of Birkhoff's theorem
for the cosmological constant is applicable \cite{Eiesland:1925,Schleich:2009uj}.\footnote{~For other generalizations, see Theorem 15.5 in \cite{Stephani:2009exact}
and references therein.} However, when $\Lambda(v)$ varies, effectively there is no vacuum
since $V_{g}(S)$ and $V_{f}(S)$ vary across the manifold and behave
as two matter stress-energy tensors. So, even in the absence of other
matter sources, whenever the bimetric potential is not constant, the
bimetric field equations are not in vacuum in the GR sense because
the two metrics act as matter sources to each other.\footnote{~Note that the bimetric potential (\ref{eq:V}) is nondynamical in
our case, $V(S)=\alpha^{2}\Lambda^{2}(v)/(3\beta_{2})$.} Therefore, the ``vacuum'' of bimetric equations is not the same
as the ``vacuum'' of GR since the latter implies absence of any
source.

\bHlineC

To illustrate this argument, consider a GR spacetime sourced by the
following matter field in the spherical null chart $x^{\mu}=(v,r,\theta,\phi)$,
\begin{align}
T_{\mu\nu}^{g} & =\left(\begin{array}{cc|cc}
\Lambda(v)\left(1-\frac{1}{3}\Lambda(v)\,r^{2}\right)+\frac{1}{3}\Lambda^{\prime}(v)r\  & \ -\Lambda(v) & 0 & 0\\
-\Lambda(v) & 0 & 0 & 0\\
\hline 0 & 0 & -\Lambda(v)\,r^{2} & 0\\
0 & 0 & 0 & -\Lambda(v)\,r^{2}\sin^{2}\theta
\end{array}\right)\!.\label{eq:se-1}
\end{align}
This stress-energy tensor has a double null eigenvector and generally
belongs to Type II fluids defined in \cite{Hawking:1973large}. As
can be shown, the (bi)metric $g$ (\ref{eq:sol-g1}) satisfies the
GR field equations $G_{\mu\nu}^{g}=T_{\mu\nu}^{g}$ for the stress-energy
tensor (\ref{eq:se-1}) since $T_{\mu\nu}^{g}=-V_{\mu\nu}^{g}$.

\eHlineC

To identify physical components of the stress-energy tensor (\ref{eq:se-1}),
it can be decomposed along the ingoing $\ell_{g}^{\mu}$ and the outgoing
$n_{g}^{\mu}$ null radial vectors of the complex null tetrad (\ref{eq:null-tetrad})
of $g$ as,
\begin{align}
T_{\mu\nu}^{g} & =\mu_{g}\,\ell_{\mu}^{g}\ell_{\nu}^{g}-\rho_{g}\,g_{\mu\nu}^{\perp}+p_{g}\,\tilde{g}_{\mu\nu}=\mu_{g}\,\ell_{\mu}^{g}\ell_{\nu}^{g}+2(\rho_{g}+p_{g})\,\ell_{(\mu}^{g}n_{\nu)}^{g}+p_{g}\,g_{\mu\nu},\label{eq:se-2}
\end{align}
where $g_{\mu\nu}^{\perp}=-2\ell_{(\mu}^{g}n_{\nu)}^{g}$ is the induced
metric on the surfaces of constant $(v,r)$ and $\tilde{g}_{\mu\nu}=g_{\mu\nu}-g_{\mu\nu}^{\perp}$
is a two-dimensional transverse metric on the normal space to those
surfaces. The components of (\ref{eq:se-2}) have a direct physical
interpretation: $\mu_{g}$ is the \emph{energy flux} in the inward
$\ell_{g}^{\mu}$ direction, $\rho_{g}$ is the \emph{energy density}
and $p_{g}$ is the \emph{tangential pressure}. Matching (\ref{eq:se-1})
with (\ref{eq:se-2}), we conclude that,
\begin{equation}
\mu_{g}=\frac{1}{3}\Lambda^{\prime}(v)\,r,\qquad\rho_{g}=\Lambda(v),\qquad p_{g}=-\Lambda(v).\label{eq:se-3}
\end{equation}
The nonvanishing flux of energy $\mu_{g}$ along $\ell_{g}^{\mu}$
classifies (\ref{eq:se-1}) as a nonperfect fluid. From the equation
of state $w_{g}=p_{g}/\rho_{g}$, we obtain $w_{g}=-1$, which is
the same as for the ordinary cosmological constant.

Hence, without any referral to bimetric theory, the metric in $g$-sector
(\ref{eq:sol-g1}) satisfies the Einstein field equations sourced
by the matter field (\ref{eq:se-1}). Obviously, this is not a GR
vacuum solution. In fact, the metric (\ref{eq:sol-g1}) has the form
of the Husain null fluid spacetimes \cite{Husain:1995bf} and the
generalized Vaidya metric \cite{Wang:1998qx}, analyzed in \cite{Ibohal:2004kk,Ibohal:2006ez,Ibohal:2009px}
in the context of nonstationary de Sitter cosmological models in GR.
In comparison to the ordinary (anti)de-Sitter spacetime with a constant
$\Lambda$, the stress-energy tensor (\ref{eq:se-1}) comprises a
nonvanishing flux of energy $\mu_{g}=\frac{1}{3}\Lambda^{\prime}(v)\,r\ne0$
in the inward direction. 

In the bimetric context, the matter source (\ref{eq:se-1}) can be
interpreted as a conformally flat nonperfect null fluid shared by
two metrics (this follows from $V_{\mu\nu}^{g}=-T_{\mu\nu}^{g}$).
In summary, for a general $\Lambda(v)$, our solution is spherically
symmetric, nonstationary (thus nonstatic) and conformally flat (Type
O by Petrov classification), sourced by an inhomogeneous nonperfect
null fluid of Type II with negative pressure.

The null energy condition for the matter distribution (\ref{eq:se-1})
requires $T_{\mu\nu}^{g}k_{g}^{\mu}k_{g}^{\nu}\ge0$ where $k_{g}^{\mu}$
is an arbitrary future-pointing null vector relative to $g$, which
can be defined by real parameters $x$, $y$ and $z$ (see appendix
B in \cite{Creelman:2016laj}),
\begin{equation}
k_{g}^{\mu}=\frac{\ee^{x}}{\sqrt{2}}\left(\ee^{y}\ell_{g}^{\mu}+\ee^{-y}n_{g}^{\mu}\right)+\frac{\ee^{x}}{\sqrt{2}}\left(\ee^{z}m_{g}^{\mu}+\ee^{-z}\bar{m}_{g}^{\mu}\right).
\end{equation}
Using (\ref{eq:null-tetrad}) we obtain $T_{\mu\nu}^{g}k_{g}^{\mu}k_{g}^{\nu}=\frac{1}{2}\ee^{2(x-y)}\mu_{g}\ge0$,
which constrains $\Lambda^{\prime}(v)\geq0$. Note also that $\rho_{g}+p_{g}=0$.
The weak energy condition states $T_{\mu\nu}^{g}u_{g}^{\mu}u_{g}^{\nu}\ge0$,
where $u_{g}^{\mu}$ is an arbitrary future-pointing timelike vector
relative to $g$. As can be shown, this requires $\rho_{g}=\Lambda(v)\ge0$
and $\mu_{g}=\frac{1}{3}\Lambda^{\prime}(v)\,r\ge0$, which constrains
$\beta_{2}\ge0$ and $\lambda^{\prime}(v)\ge0$. The strong energy
condition is always violated since $\rho_{g}$ and $p_{g}$ have opposite
signs. 

\begin{figure}
\noindent \begin{centering}
\includegraphics{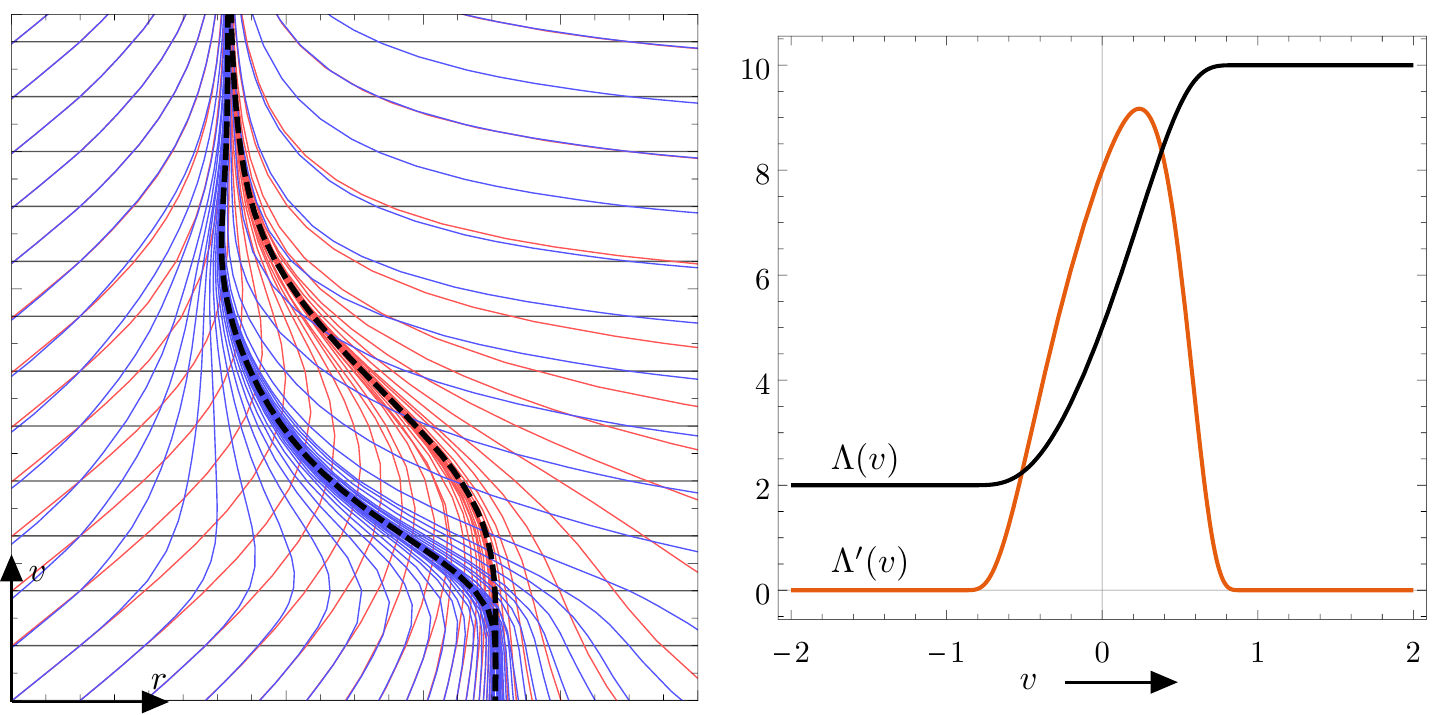}
\par\end{centering}
\caption{\label{fig:shell}Null radial geodesics (left panel) for a null shell
generated by a smooth curvature field $\Lambda(v)$ given in the right
panel. The common ingoing null radial geodesics are horizontal lines
in black. The outgoing geodesics of $g$ are depicted in red while
the outgoing geodesics of $f$ are in blue. }
\end{figure}

An example of the solution (\ref{eq:sol}) satisfying the energy conditions
$\beta_{2}\geq0$ and $\lambda^{\prime}(v)\geq0$ is shown on Figure
\ref{fig:shell}. It represents a smooth transition between two regions
of spacetime having different cosmological constants along the advanced
time direction.

Treating the $f$-sector in a similar manner, we find that the stress-energy
tensor $T_{\mu\nu}^{f}=-V_{\mu\nu}^{f}$ comprises the following physical
components,
\begin{equation}
\mu_{f}=-\alpha^{-2}\lambda^{-4}(v)\,\mu_{g},\qquad\rho_{f}=\lambda^{-2}(v)\,\rho_{g},\qquad p_{f}=\lambda^{-2}(v)\,p_{g},
\end{equation}
with the equation of state $w_{f}=p_{f}/\rho_{f}=w_{g}=-1$. The null
energy condition for the $f$-sector is similarly $T_{\mu\nu}^{f}k_{f}^{\mu}k_{f}^{\nu}=\frac{1}{2}\ee^{2(x-y)}\mu_{f}\ge0$
with $\rho_{f}+p_{f}=0$. Importantly, the signs of $\mu_{g}$ and
$\mu_{f}$ are opposite, so the null energy conditions for the two
sectors are strongly anticorrelated, in agreement with \cite{Baccetti:2012re}.
The two null energy conditions can only be simultaneously fulfilled
for $\mu_{g}=\mu_{f}=0$, which happens whenever $\Lambda(v)=\mathrm{const}$.

Note that the violation of the null energy condition is intrinsic
for the effective stress-energy tensors in the bimetric theory and
does not automatically render the solutions nonphysical \cite{Baccetti:2012re}.
Also, the usage of energy conditions in GR is motivated largely by
technical requirements about minimal assumptions needed to prove certain
theorems (e.g., the singularity theorems, the positive energy theorem,
or superluminal censorship), and many interesting GR solutions violate
some of the energy conditions \cite{Visser:1999de,Martin-Moruno:2017exc}.
In the ghost-free bimetric theory, similar considerations are more
involved and the implications of the energy conditions need further
investigation.

\section{Discussion}

\label{sec:discussion}

In this paper we employed \emph{vacuum} to denote an empty space in
the absence of nongravitational fields. This is on the line with the
majority of bimetric literature \cite{Schmidt-May:2015vnx,deRham:2014zqa}.
In GR, however, the term vacuum has a slightly different connotation.
A \emph{vacuum solution} is a spacetime where the Ricci (or Einstein)
tensor vanishes identically, $R_{\mu\nu}=0$ \cite{Stephani:2009exact,Griffiths:2009exact,Petrov:1969einstein},
which means that the stress-energy tensor also vanishes identically.
Also, an \emph{Einstein solution} is a spacetime where the Ricci tensor
is proportional to the metric, $R_{\mu\nu}\propto g_{\mu\nu}$. In
such a case, the Einstein field equations are supplemented by a cosmological
constant term (for this reason, Einstein spaces are sometimes denoted
as \emph{lambdavacuum solutions}). Hence, one has to be careful when
talking about vacuum solutions in the bimetric context while referring
to GR results. In GR, Birkhoff's theorem is strictly stated for vacuum,
$R_{\mu\nu}=0$, and can be extended to the cosmological constant
case. In bimetric theory, the cosmological constant case holds whenever
the bimetric potential $V(S)$ is constant. As shown, a similar extension
of the theorem does not work when $V(S)$ varies. 

As noted, the condition on the $\beta$-parameters (\ref{eq:PM-cond})
is the same as in the context of partially massless (PM) bimetric
gravity, so an intriguing question arises about a possible relation
between the found solutions and PM. In our case, the origin of (\ref{eq:PM-cond})
is the requirement that $\lambda(v)$ is an arbitrary function which
satisfies the equation (\ref{eq:PM-equation}). In the PM case, the
similar equation is posed for de Sitter background with a proportionality
constant $c$ between the metrics $f=c^{2}g$ \cite{Hassan:2012gz}.
The requirement for $c$ to be undetermined by the background equation
imposes the PM parameter choice. Treating the question of whether
the candidate nonlinear theory can have the full PM gauge symmetry
beyond the de Sitter background, the authors of \cite{Hassan:2012gz}
pointed out that the theory specified by (\ref{eq:PM-cond}) has an
additional nonlinear gauge symmetry for a particular nonproportional
homogeneous and isotropic vacuum solution. That solution is nonstationary
and spherically symmetric admitting six Killing vectors fields. As
in our case, the solution can be parametrized by an arbitrary function,
enabled by (\ref{eq:PM-cond}). (Since the explicit form of that solution
is not given in \cite{Hassan:2012gz}, some of its geometrical properties
are derived in appendix \ref{app:nonprop} for comparison.) Such a
similarity to \cite{Hassan:2012gz} makes our solution a suitable
choice for a non-Einstein background in the context of the candidate
PM theory. 

If the PM parameters (\ref{eq:PM-cond}) are not imposed, the equation
(\ref{eq:PM-equation}) requires $\lambda(v)=\mathrm{const}$. Then,
$\lambda$ is determined as a solution of the quartic equation (\ref{eq:PM-equation})
depending on arbitrary $\beta$-parameters. Furthermore, the integration
constant $b_{0}$ (introduced in Step~5 on page~\pageref{par:step-5})
can be chosen arbitrarily. This gives two Schwarzschild-(anti)de Sitter
metrics having different Killing horizons and cosmological constants
in general. Such solutions are encountered in \cite{Babichev:2015xha}
and considered in \cite{Torsello:2017zz} in the context of symmetries
in bimetric theory.

Another comment is about the choice of the principal square root branch
$\lambda(v)>0$ in the ansatz (\ref{eq:ansatz}). Noting that the
potential $V(S)$ in (\ref{eq:V}) is a homogeneous function of $S$,
a sign change of $\lambda$ is equivalent to the reparametrization
$\beta_{n}\to(-1)^{n}\beta_{n}$. Note also that the signs of the
$\dd v\dd r$-components in (\ref{eq:ansatz}) need not be correlated
for the two metrics as a general assumption. However, if the signs
are different, the square root (\ref{eq:mat-S}) will be purely imaginary,
in which case substituting $\lambda\to\ii\lambda$ and $G\to-G$ yields
the same equations of motion as earlier. This poses no problem if
the $\beta$-parameters satisfy (\ref{eq:PM-cond}) since $\beta_{1}$
and $\beta_{3}$ are identically zero. Nevertheless, the signature
of the metric $f$ is changed to $(+,-,-,-)$ while the signature
of $g$ remains $(-,+,+,+)$, which takes such a solution out of the
scope of this paper.

Relinquishing spherical symmetry, the solution can be made axially
symmetric using the Newman-Janis algorithm \cite{Newman:1965tw} based
on a trick with a complex coordinate transformation. The application
of the algorithm on $g$ is summarized at the end of appendix \ref{app:chart-maps}.
The resulting axially symmetric metric reads,
\begin{align}
g & =-G(v,r,\theta)\,\dd v^{2}+2\,\dd v\dd r+2a\sin^{2}\theta\,\dd r\dd\phi-2a\sin^{2}\theta\left(1-G(v,r,\theta)\right)\,\dd v\dd\phi\nonumber \\
 & \qquad+\,\rho^{2}\dd\theta^{2}+\left[\rho^{2}+a^{2}\sin^{2}\theta\left(2-G(v,r,\theta)\right)\right]\sin^{2}\theta\,\dd\phi^{2},\label{eq:axi-g}
\end{align}
where $\rho^{2}\coloneqq r^{2}+a^{2}\cos^{2}\theta$, $G(v,r,\theta)\coloneqq1-\frac{1}{3}\Lambda(v)r^{4}/\rho^{2}$,
and $a$ is a real constant that parametrizes the deviation from spherical
symmetry along $\theta$ (for $a=0$, the metric is spherically symmetric).
Constructing the square root,
\begin{align}
S & =\lambda(v)\left[\,\partial_{v}\otimes\dd v+\partial_{r}\otimes\dd r+\partial_{\theta}\otimes\dd\theta+\partial_{\phi}\otimes\dd\phi\,\right]\nonumber \\
 & \qquad+\,r\lambda^{\prime}(v)\left[\,\partial_{r}\otimes\dd v+a\sin^{2}\theta\,\partial_{r}\otimes\dd\phi\,\right].\label{eq:axi-S}
\end{align}
we obtain $f$ in a similar form as $g$ in (\ref{eq:axi-g}),
\begin{align}
f & =\lambda^{2}(v)\Bigl\{-F(v,r,\theta)\,\dd v^{2}+2\,\dd v\dd r+2a\sin^{2}\theta\,\dd r\dd\phi-2a\sin^{2}\theta\left(1-F(v,r,\theta)\right)\,\dd v\dd\phi\nonumber \\
 & \qquad+\,\rho^{2}\dd\theta^{2}+\left[\rho^{2}+a^{2}\sin^{2}\theta\left(2-F(v,r,\theta)\right)\right]\sin^{2}\theta\,\dd\phi^{2}\Bigl\},\label{eq:axi-f}
\end{align}
where $F(v,r,\theta)=G(v,r,\theta)+2r\lambda^{\prime}(v)/\lambda(v)$.
The square root (\ref{eq:axi-S}) differs from (\ref{eq:sol-S}) in
the presence of a new off-diagonal component $\tud Sr{\phi}=ar\sin^{2}\theta\lambda^{\prime}(v)$,
which will contribute to the effective stress-energy $V_{g}$ with
the off-diagonal component,
\begin{equation}
\tud{V_{g}}r{\phi}=-\frac{1}{3}\Lambda^{\prime}(v)r\,a\sin^{2}\theta.
\end{equation}
The Einstein tensors of $g$ and $f$ will have complicated forms
(the explicit form of $G_{g}$ can be found in \cite{Ibohal:2004kk}),
with the resulting $G_{g}+V_{g}$ and $G_{f}+V_{f}$ being nonzero.
These can be matched with additional stress-energy tensors having
off-diagonal components with nonvanishing rotation as a part of the
modified energy flux, energy density and pressure.

Finally, we comment on the relation to GR solutions. The bimetric
potential $V(S)$ of the found solution is a nondynamical field that
is not governed by equations of motion. Thus the two sectors are dynamically
decoupled, similarly to a bi-Einstein space setup with a constant
$V(S)$ \cite{Hassan:2014vja,Kocic:2017wwf}. Here, however, we have
two GR sectors each sourced by its own stress-energy tensor, where
the two stress-energy tensors are related by (\ref{eq:V-identity}).
For our solution, the stress-energy tensors are of the generalized
Vaidya type, as noted in subsection \ref{sec:physics}. In GR, the
Vaidya metric is a solution of the Einstein field equations describing
the spacetime of a spherically symmetric inhomogeneous imploding (exploding)
null dust fluid \cite{Vaidya:1951zza}. It is a nonstatic generalization
of the Schwarzschild solution and admits only three independent Killing
vector fields. The Vaidya metric is given by (\ref{eq:ansatz-g})
with $G(v,r)=1-2m(v)/r$ and $p=0$, where $m(v)$ is called the \emph{mass
function}, related to the gravitational energy within a given radius
$r$ \cite{Lake:1991bff,Poisson:1990eh}. Note also that, unlike the
Vaidya metric, our solution does not have a $1/r$ term. The mass
function $m(v)$ can be generalized to depend also on the radial coordinate
$r$ as in \cite{Husain:1995bf} and \cite{Wang:1998qx}. In the latter
work, the mass function was considered to be expanded in the powers
of $r$ as,
\begin{equation}
m(v,r)=\sum_{n=-\infty}^{\infty}a_{n}(v)\,r^{n},
\end{equation}
where $a_{n}(v)$ are arbitrary functions depending only on $v$.
Such generalized Vaidya solutions were further analyzed in \cite{Ibohal:2004kk,Ibohal:2006ez,Ibohal:2009px}
in the context of nonstationary de Sitter cosmological models (both
spherically and axially symmetric). One particular model considered
a mass function with $a_{3}(v)=\frac{1}{6}\Lambda(v)$ and $a_{n\ne3}(v)=0$,
which coincides with our solution (\ref{eq:sol-g1}). In bimetric
theory, however, such a model naturally comes out with a more specific
form of $\Lambda(v)$ (\ref{eq:cosmo-field}) whenever two spherically
symmetric metrics share a common null direction in empty space.

\acknowledgments

We would like to thank Fawad Hassan, Mikael von Strauss and Angnis
Schmidt-May for helpful discussions on the relations to PM symmetry.
We are grateful to Luis Apolo for a careful reading of the manuscript.

\clearpage
\phantomsection
\addcontentsline{toc}{section}{Appendices}
\section*{Appendices}

\def\toclevel@section{1}
\def\toclevel@subsection{2}
\addtocontents{toc}{\string\let\string\l@section\string\l@subsection}

\appendix

\section{Chart transition maps}

\label{app:chart-maps}

In this appendix, we provide three chart transition maps; the first
diagonalizes the metric $g$, the second puts the metric $g$ into
a conformally flat form, and the third is a complex coordinate transformation
trick which generates an axially symmetric from a spherically symmetric
metric.

\paragraph*{Chart 1.}

Consider a chart transition map from $(v,r,\theta,\phi)$ to $(T,R,\theta,\phi)$
that diagonalizes $g$, so that,
\begin{equation}
g=-a(T,R)\,\dd T^{2}+b(T,R)\,\dd R^{2}+r^{2}\dd\Omega^{2}.\label{eq:g-TR}
\end{equation}
Assuming that the coordinate transformation is achieved through functions
$T(v,r)$ and $R(v,r)$, we have,\bSe\label{eq:TR-map}
\begin{align}
\dd T & =\partial_{v}T(v,r)\,\dd v+\partial_{r}T(v,r)\,\dd r,\\
\dd R & =\partial_{v}R(v,r)\,\dd v+\partial_{r}R(v,r)\,\dd r.
\end{align}
\eSe Substituting (\ref{eq:TR-map}) in (\ref{eq:g-TR}), then equating
with (\ref{eq:sol-g}), gives the system of partial differential equations,\bSe
\begin{align}
0 & =1-\frac{1}{3}\Lambda(v)\,r^{2}+b(v,r)\left(\partial_{v}R(v,r)\right)^{2}-a(v,r)\left(\partial_{v}T(v,r)\right)^{2},\label{eq:le-2}\\
0 & =1-2b(v,r)\,\partial_{r}R(v,r)\,\partial_{v}R(v,r)+2a(v,r)\,\partial_{r}T(v,r)\,\partial_{v}T(v,r),\\
0 & =b(v,r)\left(\partial_{r}R(v,r)\right)^{2}-a(v,r)\left(\partial_{r}T(v,r)\right)^{2}.
\end{align}
\eSe From the last two equations we can solve for $a(v,r)$ and $b(v,r)$
in terms of $\partial_{r}T(v,r)$, $\partial_{r}R(v,r)$ and the Jacobian
$J=\partial_{v}T(v,r)\partial_{r}R(v,r)-\partial_{r}T(v,r)\partial_{v}R(v,r)$,
\begin{equation}
a(v,r)=-\frac{\partial_{r}R(v,r)}{J\,\partial_{r}T(v,r)},\qquad b(v,r)=-\frac{\partial_{r}T(v,r)}{J\,\partial_{r}R(v,r)}.
\end{equation}
Substituting $a(v,r)$ and $b(v,r)$ in (\ref{eq:le-2}) yields,
\begin{equation}
\frac{1}{3}\Lambda(v)\,r^{2}=1+\frac{\partial_{v}R(v,r)}{\partial_{r}R(v,r)}+\frac{\partial_{v}T(v,r)}{\partial_{r}T(v,r)}.
\end{equation}
Note that the variables cannot be separated because of the mixed term
$\Lambda(v)r^{2}$. Nonetheless, the above equation can have many
solutions; one is easily obtained by splitting the right hand side
in two parts,
\begin{equation}
1+\frac{\partial_{v}R(v,r)}{\partial_{r}R(v,r)}=0,\qquad\frac{\partial_{v}T(v,r)}{\partial_{r}T(v,r)}=\frac{1}{3}\Lambda(v)\,r^{2}.
\end{equation}
This gives (with integration constants suppressed),
\begin{equation}
R(v,r)=r-v,\qquad T(v,r)=-\frac{1}{r}+\beta_{2}\intop\left(\alpha^{-2}+\lambda^{2}(v)\right)\dd v,\label{eq:TR-map2}
\end{equation}
which yields,
\begin{equation}
a(v,r)=\frac{-r^{2}}{r^{-2}+\beta_{2}\left(\alpha^{-2}+\lambda^{2}(v)\right)},\qquad b(v,r)=\frac{-r^{-2}}{r^{-2}+\beta_{2}\left(\alpha^{-2}+\lambda^{2}(v)\right)}.
\end{equation}
As a final step, we express $a(r,v)$ and $b(r,v)$ in terms of $T$
and $R$ by inverting (\ref{eq:TR-map2}), which is a nontrivial task
highly dependent on the form of $\lambda(v)$. Note that the chart
transition (\ref{eq:TR-map2}) is valid for the nonvanishing Jacobian
determinant $r^{-2}+\beta_{2}\left(\alpha^{-2}+\lambda^{2}(v)\right)$
(note that $r=0$ is not part of the spherical chart).

To avoid inverting (\ref{eq:TR-map2}), a more direct chart transition
can be devised by fixing $a$, $b$ and $R$ in (\ref{eq:g-TR}) as,
\begin{equation}
a(T,R)\coloneqq\varphi^{2}(T,R)\,G(T,R),\quad b(T,R)\coloneqq G(T,R)^{-1},\quad R(v,r)\coloneqq r,
\end{equation}
so that $G(T,R)=\left(1-\frac{1}{3}\Lambda(t,r)\,r^{2}\right)^{-1}$.
Here, we used the freedom to express $\lambda(v)$ and $\Lambda(v)$
in terms of a new arbitrary field $\lambda(T,R)$. Plugging in (\ref{eq:TR-map})
into (\ref{eq:g-TR}), then equating with (\ref{eq:sol-g}), we obtain
the conditions,
\begin{equation}
\frac{\partial_{v}T(v,r)}{\partial_{r}T(v,r)}+G(v,r)=0,\qquad\varphi(v,r)=\frac{1}{G(v,r)\,\partial_{r}T(v,r)}.
\end{equation}
Replacing the coordinate scalar $T$ by $t$ in terms of $\lambda(t,r)$
so that, 
\begin{equation}
\frac{1}{\partial_{r}T(v,r)}\dd T\coloneqq\frac{\partial_{t}\lambda(t,r)}{\partial_{r}\lambda(t,r)}\dd t,
\end{equation}
and further setting $\dd\lambda=\partial_{t}\lambda(t,r)\,\dd t+\partial_{r}\lambda(t,r)\,\dd r$,
we have,
\begin{equation}
\dd v\coloneqq\frac{\dd\lambda}{G(t,r)\,\partial_{r}\lambda(t,r)}=\frac{1}{G(t,r)}\frac{\partial_{t}\lambda(t,r)}{\partial_{r}\lambda(t,r)}\dd t+\frac{1}{G(t,r)}\dd r.
\end{equation}
Substituting the above $\dd v$ in (\ref{eq:sol-g1}) yields a familiar
form of the metric in the spherical chart $(t,r,\theta,\phi)$,
\begin{equation}
g=-\varphi^{2}(t,r)\left(1-\frac{1}{3}\Lambda(t,r)\,r^{2}\right)\dd t^{2}+\left(1-\frac{1}{3}\Lambda(t,r)\,r^{2}\right)^{-1}\dd r^{2}+r^{2}\dd\Omega^{2},
\end{equation}
where $\dd\Omega^{2}=\dd\theta^{2}+\sin^{2}\theta\,\dd\phi^{2}$,
$\Lambda(t,r)=3\beta_{2}\left(\alpha^{-2}+\lambda^{2}(t,r)\right)$
and,
\begin{equation}
\varphi(t,r)=\frac{\partial_{t}\lambda(t,r)}{\partial_{r}\lambda(t,r)}\left(1-\frac{1}{3}\Lambda(t,r)\,r^{2}\right)^{-1}.
\end{equation}

\paragraph*{Chart 2.}

In the following, we provide a procedure to find a chart transition
map which puts the metric $g$ into a conformally flat form,
\begin{equation}
g=\psi^{2}(T,R)\,\left[-\dd T^{2}+\dd R^{2}+R^{2}\dd\Omega^{2}\right].\label{eq:g-conf-map}
\end{equation}
The coordinate transformation is achieved through functions $T(v,r)$
and $R(v,r)$ using (\ref{eq:TR-map}). Plugging in (\ref{eq:TR-map})
into (\ref{eq:g-conf-map}), then equating with (\ref{eq:sol-g}),
we obtain the condition $\psi(v,r)=-r/R(v,r)$ and the following system
of partial differential equations,\bSe
\begin{align}
0 & =1-\frac{1}{3}\Lambda(v)\,r^{2}+\frac{r^{2}}{R^{2}(v,r)}\left[\left(\partial_{v}R(v,r)\right)^{2}-\left(\partial_{v}T(v,r)\right)^{2}\right],\\
0 & =R^{2}(v,r)-r^{2}\left[\partial_{r}R(v,r)\,\partial_{v}R(v,r)-\partial_{r}T(v,r)\,\partial_{v}T(v,r)\right],\\
0 & =\left(\partial_{r}R(v,r)\right)^{2}-\left(\partial_{r}T(v,r)\right)^{2}.\label{eq:cf-3}
\end{align}
\eSe Introducing new coordinate scalars $U$ and $V$ through 
\begin{equation}
T(v,r)=U(v,r)-V(v,r),\qquad R(v,r)=U(v,r)+V(v,r),\label{eq:cf-9}
\end{equation}
from (\ref{eq:cf-3}) we get $\partial_{r}U(v,r)\,\partial_{r}V(v,r)=0$.
As a consequence, either $U$, $V$ or both do not depend on $r$.
Setting $V(v,r)=V(v)$ yields,\bSe
\begin{align}
0 & =\left(1-\frac{1}{3}\Lambda(v)\,r^{2}\right)\left(U(v,r)+V(v)\right)^{2}+4r^{2}U(v,r)\,V^{\prime}(v),\label{eq:cf-1}\\
0 & =\left(U(v,r)+V(v)\right)^{2}-2r^{2}V^{\prime}(v)\,\partial_{r}U(v,r).\label{eq:cf-2}
\end{align}
\eSe From (\ref{eq:cf-2}) we can solve,
\begin{equation}
U(v,r)=\frac{2r\,V^{\prime}(v)}{1-r\,W^{\prime}(v)\,V^{\prime}(v)}-V(v),\label{eq:cf-8}
\end{equation}
where $W(v)$ is an arbitrary integration function. Substituting $U(v,r)$
in (\ref{eq:cf-1}), then expanding as a series in $r$, yields the
system,\bSe
\begin{align}
0 & =W^{\prime}(v)V^{\prime}(v)+\frac{V^{\prime\prime}(v)}{V^{\prime}(v)},\label{eq:cf-4}\\
\frac{1}{3}\Lambda(v) & =\left[2W^{\prime\prime}(v)-W^{\prime}(v)^{2}V^{\prime}(v)\right]V^{\prime}(v).\label{eq:cf-5}
\end{align}
\eSe From (\ref{eq:cf-4}) we obtain,
\begin{equation}
V^{\prime}(v)=\frac{1}{W(v)+c_{1}},\qquad V(v)=\int V^{\prime}(v)\,\dd v+c_{2},\label{eq:cf-6}
\end{equation}
where $c_{1}$ and $c_{2}$ are integration constants. Substituting
in (\ref{eq:cf-5}) gives,
\begin{equation}
2W^{\prime\prime}(v)\,\left[W(v)-c_{1}\right]-W^{\prime}(v)^{2}-\frac{1}{3}\Lambda(v)\left[W(v)-c_{1}\right]^{2}=0.\label{eq:cf-7}
\end{equation}
This equation can be solved for $W$ depending on the form of $\Lambda(v)$.
Using $W$ from (\ref{eq:cf-7}), $V$ from (\ref{eq:cf-6}), and
$U$ from (\ref{eq:cf-8}) yields,
\begin{equation}
\varphi(v,r)=\frac{1}{2}\left(r\,W^{\prime}(v)-\frac{1}{V^{\prime}(v)}\right)\!,\qquad R(v,r)=\frac{2r\,V^{\prime}(v)}{1-r\,W^{\prime}(v)\,V^{\prime}(v)},
\end{equation}
where $T(v,r)=R(v,r)-2V(v)$. As a final step, one should express
$\varphi(v,r)$ in terms of $T$ and $R$ by inverting (\ref{eq:TR-map2}),
which nontrivally depends on the form of $\Lambda(v)$.

\paragraph*{Generating an axially symmetric metric.}

Here we follow the algorithm from \cite{Newman:1965tw} to ``derive''
an axially symmetric metric from a known spherically symmetric one
(see also \cite{Newman:1965my} and section 2 in \cite{Ibohal:2004kk}).
Consider a spherically symmetric metric of the form (\ref{eq:sol-g1}),
\begin{align}
g & =-G\,\dd v^{2}+2\,\dd v\dd r+r^{2}\left(\dd\theta^{2}+\sin^{2}\theta\,\dd\phi^{2}\right),
\end{align}
where $G=1-\frac{1}{3}\Lambda(v)r^{2}$. Allowing $r$ to take complex
values (so that $\bar{r}$ is the complex conjugate of $r$), we formally
perform the complex coordinate transformation,
\begin{equation}
V=v+\ii a\cos\theta,\qquad R=r+\ii a\cos\theta,\label{eq:le-3}
\end{equation}
where $V$ and $R$ are considered to be real. Here, $a$ is a real
constant that parametrizes the deviation from the spherical symmetry.
Using the map (\ref{eq:le-3}) and a suitable substitution $\dd\theta\leftrightarrow-\ii\sin\theta\dd\phi$
(prescribed in \cite{Ibohal:2004kk}), we get,
\begin{align}
g & =-G\,\dd V^{2}+2\,\dd V\dd R+2a\sin^{2}\theta\,\dd R\dd\phi+2a\sin^{2}\theta\left(1-G\right)\,\dd V\dd\phi\nonumber \\
 & \qquad+\,\left(R^{2}+a^{2}\cos^{2}\theta\right)^{2}\dd\theta^{2}+\left[R^{2}+a^{2}+a^{2}\sin^{2}\theta\left(1-G\right)\right]\sin^{2}\theta\,\dd\phi^{2},
\end{align}
where $G=1-\frac{1}{3}\Lambda(v)R^{4}/\left(R^{2}+a^{2}\cos^{2}\theta\right)$. 

Note that the chart $(V,R,\theta,\phi)$ has a coordinate singularity
at $R^{2}+a^{2}\cos^{2}\theta=0$. Finally, after conveniently replacing
$V\to v$ and $R\to r$, we obtain (\ref{eq:axi-g}).

\section{Nonproportional vacuum solution}

\label{app:nonprop}

To obtain a solution parametrized by an arbitrary function, the authors
of \cite{Hassan:2012gz} consider nonproportional homogeneous and
isotropic backgrounds \cite{vonStrauss:2011mq},
\begin{equation}
g=-\dd t^{2}+a^{2}(t)\,h,\qquad f=-X^{2}(t)\dd t^{2}+Y^{2}(t)\,h,\label{eq:hom-iso-gf}
\end{equation}
where $h$ is the spatial metric with the curvature $k$,
\begin{equation}
h=\frac{\dd r^{2}}{1-kr^{2}}+r^{2}\left(\dd\theta^{2}+\sin^{2}\theta\,\dd\phi^{2}\right).
\end{equation}
Here, $a(t)$, $Y(t)$ and $X(t)$ are three fields that parametrize
the ansatz. In general, these functions can be solved for from the
bimetric field equations \cite{vonStrauss:2011mq}. In vacuum, the
equation that determines $Y/a$ is identically zero provided the PM
parameter choice (\ref{eq:PM-cond}) is satisfied. Consequently, one
of the three functions in (\ref{eq:hom-iso-gf}) is arbitrary. Note
that this ansatz does not allow for an explicit stress-energy tensor
in either $g$ or $f$, but does allow for it if both $g$ and $f$
have seperate $T_{\mu\nu}$ (similarly to the axially symmetric case).
In this case however, the ratio $Y/a$ will be determined by an equation
of motion. 

For comparison to our solution, consider the equivalent ansatz,
\begin{align}
g & =-\dd t^{2}+\ee^{2a(t)}\,h,\qquad f=\lambda^{2}(t)\left[-X^{2}(t)\dd t^{2}+\ee^{2a(t)}\,h\right]\!.
\end{align}
Here, the field $\lambda(t)$ corresponds to $Y/a$ in (\ref{eq:hom-iso-gf}).
Also, the field $a(t)$ is replaced by $\ee^{a(t)}$ to simplify equations.
The bimetric field equations are,\bSe
\begin{align}
0 & =\left\langle \lambda\right\rangle _{0}^{3}-3a^{\prime2},\\
0 & =\left\langle \lambda\right\rangle _{0}^{2}+X\left\langle \lambda\right\rangle _{1}^{2}-2a^{\prime\prime}-3a^{\prime2},\\
0 & =\alpha^{-2}X^{2}\left\langle \lambda\right\rangle _{1}^{3}-3\lambda\left(\lambda a^{\prime}+\lambda^{\prime}\right)^{2},\\
0 & =\alpha^{-2}X^{2}\left(\langle\left\langle \lambda\right\rangle _{1}^{2}+X\langle\lambda\rangle_{2}^{2}\right)+2\lambda X^{\prime}\left(\lambda a^{\prime}+\lambda^{\prime}\right),\\
 & \qquad-X\left[2\lambda\left(3a^{\prime}\lambda^{\prime}+\lambda^{\prime}\right)+\lambda^{2}\left(2a^{\prime\prime}+3a^{\prime2}\right)+\lambda^{\prime2}\right].
\end{align}
\eSe where derivatives are with respect to $t$. The Bianchi constraint
is,
\begin{equation}
\left\langle \lambda\right\rangle _{1}^{2}\left[\lambda^{\prime}+\left(\lambda-X\right)a^{\prime}\right]=0.
\end{equation}
From the Bianchi constraint we have $X(t)=\lambda(t)+\lambda^{\prime}(t)/a^{\prime}(t)$.
Then for the PM parameter choice (\ref{eq:PM-cond}), we obtain the
solution,
\begin{equation}
a^{\prime}(t)^{2}=\beta_{2}\left(\alpha^{-2}+\lambda^{2}(t)\right),
\end{equation}
where $\lambda(t)$ is an arbitrary function. Similarly to (\ref{eq:cosmo-field}),
we can define,
\begin{equation}
\Lambda(t)\coloneqq3\beta_{2}\left(\alpha^{-2}+\lambda^{2}(t)\right).
\end{equation}
Then $\Lambda(t)=3a^{\prime2}(t)$ and $a(t)=\int\!\left(\Lambda(t)/3\right)^{1/2}\dd t$.
The resulting square root is in matrix notation,
\begin{equation}
S=\diag\left[\,\lambda(t)+\lambda^{\prime}(t)/a^{\prime}(t),\,\lambda(t),\lambda(t),\lambda(t)\,\right].
\end{equation}
This gives the effective stress energy tensor $T_{\mu\nu}^{g}=-V_{\mu\nu}^{g}$
with the nonzero components,
\begin{equation}
\rho_{g}=T_{00}^{g}=\Lambda(t),\qquad p_{g}=T_{11}^{g}=-\left(\Lambda(t)+\frac{\Lambda^{\prime}(t)}{\sqrt{3\Lambda(t)}}\right)\exp\left[2\int\!\left(\Lambda(t)/3\right)^{1/2}\dd t\right].
\end{equation}
From the equation of state,
\begin{equation}
w_{g}=\frac{p_{g}}{\rho_{g}}=-\left(1+\frac{\Lambda^{\prime}(t)}{\sqrt{3}}\Lambda^{-3/2}(t)\right)\exp\left[2\int\!\left(\Lambda(t)/3\right)^{1/2}\dd t\right].
\end{equation}
Note that $\Lambda(t)$ is arbitrary and can be obtained from observations.
Adjusting $\Lambda(t)$ can also be used to model inflationary scenarios.
The solution is homogeneous and isotropic with six Killing vector
fields. The chart map $\dd\tilde{t}=\lambda(t)X(t)\dd t$ puts the
metric $f$ in a similar form as $g$, relating the Killing vector
fields of the two sectors.

\ifprstyle

\bibliographystyle{apsrev4-1}
\bibliography{ns-ss-vs}

\else

\bibliographystyle{JHEP}
\bibliography{ns-ss-vs}

\fi
\end{document}